\documentclass[structabstract]{aa}

\usepackage{graphicx}
\usepackage{txfonts}
\usepackage{tabularx}

\begin{document}

\title{The spatial extent of Polycyclic Aromatic Hydrocarbons emission in the Herbig star HD\,179218}
\subtitle{} 
\titlerunning{Extent of PAHs emission in HD\,179218}

\author{A. S. Taha \inst{1,2}
\and L. Labadie\inst{1}
\and E. Pantin\inst{3}
\and A. Matter\inst{4}
\and C. Alvarez \inst{5}
\and P. Esquej\inst{6}
\and R. Grellmann\inst{1}
\and \\R. Rebolo\inst{7}
\and C. Telesco\inst{8}
\and S. Wolf\inst{9}
}

\institute{I. Physikalisches Institut, Universit\"at zu K\"oln, Z\"ulpicher Stra\ss e 77, 50937 K\"oln, Germany\\
\email{maotahni@ph1.uni-koeln.de,labadie@ph1.uni-koeln.de}
\and Department of Astronomy, Faculty of Science, University of Baghdad, Baghdad - Aljadirya, Iraq
\and Laboratoire AIM, CEA/DRF--CNRS--Universit\'e Paris Diderot, IRFU/DAS, F-91191 Gif-sur-Yvette, France
\and Laboratoire Lagrange, CNRS UMR 7293, UNS -- Observatoire de la C\^ote d'Azur, BP 4229, 06304 Nice Cedex 4, France
\and W. M. Keck Observatory, 65-1120 Mamalahoa Hwy, Kamuela, HI 96743
\and European Space Astronomy Centre (ESAC)/ESA, P.O. Box 78, 28690 Villanueva de la Ca\~nada, Madrid, Spain 
\and Instituto de Astrofisica de Canarias, C/ Via Lactea s/n, La Laguna, 38200 Tenerife , Spain
\and Department of Astronomy, University of Florida, Gainesville, FL 32611, USA
\and Institute for Theoretical Physics and Astrophysics, University of Kiel, Leibnizstr. 15, 24118 Kiel, Germany
}

\date{Received 04 07 2013/ Accepted xx xxx xxxx}

\abstract{
{\it Aims.} We investigate, in the mid-infrared, the spatial properties of the 
Polycyclic Aromatic Hydrocarbons (PAHs) emission in the disk of HD\,179218, an intermediate-mass Herbig star at $\sim$300\,pc.\\
{\it Methods.} We obtained mid-infrared images in the PAH-1, PAH-2 and Si-6 filters centered at 8.6, 11.3, and 12.5\,$\mu$m, and N-band low-resolution spectra using CanariCam on the 10-m Gran Telescopio Canarias (GTC). We compared the point spread function (PSF) 
profiles measured in the PAH filters to the profile derived in the Si-6 filter, where the thermal continuum emission dominates. We performed radiative transfer modeling of the spectral energy distribution and produced synthetic images in the three filters to investigate different spatial scenarios. \\
{\it Results.} Our data show that the disk emission is spatially resolved in the PAH-1 and PAH-2 filters, while unresolved in the Si-6 filter. Thanks to very good observing conditions, an average full width at half maximum (FHWM) of 0.232$^{\prime\prime}$, 0.280$^{\prime\prime}$ and 0.293$^{\prime\prime}$ is measured in the three filters, respectively. Gaussian disk fitting and quadratic subtraction of the science and calibrator PSFs suggests a lower-limit characteristic angular diameter of the emission of $\sim$100\,mas, or $\sim$40\,au. 
The photometric and spectroscopic results are compatible with previous findings. 
Our radiative transfer (RT) modeling of the continuum suggests that the resolved emission should result from PAH molecules on the disk atmosphere being UV-excited by the central star. Simple geometrical models of the PAH component compared to the underlying continuum point to a PAH emission uniformly extended out to the physical limits of the disk model. Furthermore, our RT best model of the continuum requires a negative exponent of the surface density power-law, in contrast with earlier modeling pointing to a positive exponent.\\
{\it Conclusions.} We have spatially resolved -- for the first time to our knowledge -- the PAHs emission in the disk of HD\,179218 and set constraints on its spatial extent. Based on spatial and spectroscopic considerations as well as on qualitative comparison with IRS\,48 and HD\,97048, we favor a scenario in which PAHs extend out to large radii across the flared disk surface and are at the same time predominantly in an ionized charge state due to the strong UV radiation field of the 180$L_{\odot}$ central star. 
}  
 

\maketitle

\section{Introduction}

\begin{table*}[t]
\centering
\caption{Observing log. Frames in the PAH-1 and PAH-2 filters were acquired on June 8$^{\rm }$, 2015. Frames in the Si-6 filter were acquired on September 15$^{\rm }$, 2015.}
\label{Table_1} 
\begin{tabular}{c c c c c c c c c}       
\hline\hline
Object & Start[UT]& End[UT] & Filter &On source [Sec] & PWV [mm] & Airmass & Seeing["]& Flux standard \\[0.4ex] 
  \hline
HD\,169414 & 03:35 & 03:46 & PAH-1 & 397 & 7.8 & 1.04 & 0.60& PSF standard star\\
HD\,179218 & 04:01 & 04:49 & PAH-1 & 1853 & 8.1 &1.06 & 0.42$\pm$0.03 & Science\\
HD\,187642 & 05:37 & 05:48 & PAH-1 & 397 & 7.6 &1.17 &0.55 & PSF standard star \\
 \hline
HD\,169414 & 03:47 & 03:57 &PAH-2 &417 & 7.7 & 1.06 &  0.51&PSF standard star \\ 
HD\,179218 & 04:51 & 05:33 & PAH-2&1807 & 8.1 &1.14 &  0.45$\pm$0.02 & Science \\
HD\,187642 & 05:48 & 05:58 & PAH-2 &417 & 7.0 &1.20 &0.58& PSF standard star \\
\hline
HD\,169414 & 20:55 & 21:06 &Si-6 &397 & 8 & 1.03 &  1&PSF standard star \\ 
HD\,179218 & 21:40 & 22:28 & Si-6 &1853 & 7.7 &1.08 &  0.60$\pm$0.03 & Science \\
HD\,187642 & 22:34 & 22:44 & Si-6 &397 & 8.2 &1.11 &0.5& PSF standard star \\
 \hline
HD\,169414 & 00:38 & 00:49 &LR Spec. & 176 & 11.3 & 1.0 &  0.85 & Telluric standard  \\ 
HD\,179218 & 01:13 & 02:11 & LR Spec. & 943 & 11.1 & 1.02 &  0.77 & Science \\
HD\,187642 & 02:32 & 02:43 & LR Spec. & 176 & 10 &1.06 &0.80& Telluric standard \\
 \hline
\end{tabular}
\end{table*}

Circumstellar disks around pre-main sequence stars constitute the reservoir of gas and dust out of which planetary systems may form. The study of their morphological structure and spectroscopic content, as well as of their temporal evolution provides important information used to constrain the models of planet formation. 
From the spectral shape of the infrared excess, \cite{Meeus2001} classify intermediate-mass Herbig stars as two groups based on the possible geometry of their dust disks. Group\,I objects exhibit a flared disk geometry while group\,II sources correspond to a flatter geometry of the circumstellar disk. 
Recently, a number of high-angular-resolution and high-sensitivity spectroscopic studies have provided evidence showing the complex spatial structure of disks in the form of (pre-)transitional or "gaped" disks eventually harboring spiral structures 
\citep{Calvet2002, Furlan2006, Espaillat2010, Graefe2011, Tatulli2011, Benisty2015}. 
A powerful tracer of the possible flared structure of the disk is found in the Polycyclic aromatic hydrocarbons (PAHs) mid-infrared emission bands found in a significant number of Herbig stars \citep{Acke2010}. 
When in the direct line-of-sight of the central star, PAH molecules on the surface of a flared disk can be electronically UV-excited by stellar photons even at large distances in the disk and cool down by re-emitting in the CH- or CC- stretching and bending modes at characteristic wavelengths (e.g., at 6.3, 8.6, or 11.3$\mu$m). High-spatial-resolution imaging and long-slit spectroscopy in the PAH bands has exploited these properties to investigate the outer disk structure in HD\,97048 \citep{Lagage2006}, or to trace possible gas flows through the disk's gaps (Maaskant et al. 2014). PAHs emission also trace the presence of very small grains mixed with the gas at high elevation above the midplane and have a significant influence on the structure of the disk by contributing to the gas heating \citep{Habart2004c}.\vspace{0.20cm} \newline 
In this paper we present CanariCam \citep{Telesco2003} high-angular-resolution mid-infrared imaging and spectroscopy data for HD\,179218, a Herbig star located at $\sim$290\,pc\footnote{New GAIA parallax, see Sect.~\ref{zdiskmodel}} 
with a B9 spectral type that harbors a circumstellar disk primarily revealed through its infrared excess. \cite{Meeus2001} classified 
HD\,179218 as a group-Ia source, which suggests a flared disk structure. 
A large amount of crystalline grains are found in this source, which points to significant dust processing \citep{Bouwman2001}. 
The latter two papers report the detection of PAHs at 8.6\,$\mu$m and at 11.3\,$\mu$m. 
These detections were later confirmed and quantified by \cite{Juhasz2010}. 
Regarding the spatial structure of HD\,179218's disk, \cite{Fedele2008}  used MIDI/VLTI mid-infrared interferometry to show that HD\,179218 could be a pre-transitional disk. 
Furthermore, the authors noticed a lower visibility shortward of 9\,$\mu$m that may result from a larger size scale of the PAH emission with respect to the continuum, but the scenario remains speculative based on the quality of the MIDI data.\newline
Here, we aim at resolving the disk emission in two PAH bands in order to constrain the global structure of the disk on the large scale \citep{Wolf2012} as we may assume that the PAH molecules remain co-spatial with the gas \citep{Woitke2016}. 
The paper is structured as follows: Section~2 summarizes the new observations conducted on the GTC. Section~3 presents the observational imaging and spectroscopic results and Section~4 focuses on the derivation of the emission characteristic size. Section~5 presents our modeling to investigate the origin of the resolved emission, while our results are discussed in Section~6.
   
\section{Observations and data reduction}\label{sec:obs_red}

We used CanariCam \citep{Telesco2003}, the mid-infrared (7.5 - 25 $\mu m$) imager with spectroscopic capabilities of the Gran Telescopio CANARIAS in La Palma, Spain. Although the GTC has an equivalent 10.4-m primary mirror, the diffraction limit achievable with CanariCam is set by a cold circular pupil stop equivalent to 9.4-m used to optimize the sensitivity. 
CanariCam holds a Raytheon 320x240 Si:As detector which covers a field of view of 26$^{\prime\prime}$$\times$19$^{\prime\prime}$. The detector plate scale is 0.08$^{\prime\prime}$/pixel\footnote{cf. http://www.gtc.iac.es/instruments/canaricam/canaricam.php}.
\\
Images were taken in the PAH-1 (8.6\,$\mu$m, $\Delta\lambda$$\approx$0.43\,$\mu$m), PAH-2 (11.3\,$\mu$m, $\Delta\lambda$$\approx$0.60\,$\mu$m) and Si-6 (12.5\,$\mu$m, $\Delta\lambda$$\approx$0.7\,$\mu$m) filters. A standard chop-nod procedure with 8$^{\prime\prime}$ throw was used. Chopping is hence performed within the detector, which means that one central positive image of the source flanked by two negative images is obtained after data reduction. 
The observations of HD\,179218 have been obtained on June 8$^{\rm }$, 2015, and September 15$^{\rm }$, 2015. 
The observing log is detailed in Table~\ref{Table_1}. 
The observation sequence consisted in the observation of a PSF reference star, followed by the science object, followed again by a second and different PSF reference star. 
The positions in the sky of the calibrators were chosen to minimize the difference in the parallactic angles at the time of observation. This ensures a similar image quality for the science and calibration targets. 
\begin{figure}[b]
\includegraphics[width=\columnwidth]{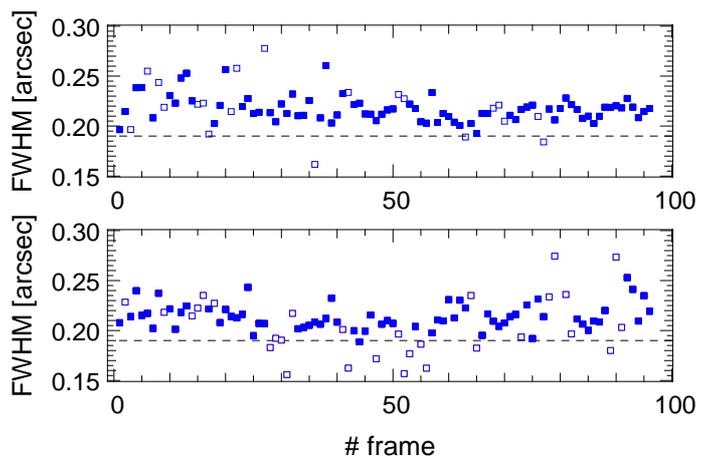}
\caption{Illustration of the frame selection on the PSF calibrators HD\,187642 (top) and HD\,169414 (bottom) for the PAH-1 filter. The full distribution is given by the filled\,+\,empty symbols. The filled squares correspond to the frames finally selected. The dashed line is the theoretical diffraction limit of the telescope at 8.6\,$\mu$m. The FWHM is estimated by fitting of a Lorentzian function.}\label{frameselection}
\end{figure}
We have performed this technique of observation for all three filters mentioned above. 
The average optical seeing extracted from the DIMM measurements was excellent during the June run, with values raging from 0.42$^{\prime\prime}$ to 0.6$^{\prime\prime}$. For the September run, the first calibrator suffered from poorer optical seeing ($\gtrsim$1$^{\prime\prime}$) in comparison to the following science target and second calibrator, which were observed in much better seeing conditions. The precipitable water vapor was measured between 7\,mm and 8\,mm, which is suitable for good-quality observations in the N band. 
Low-resolution spectroscopic observations were obtained on July 1,$^{}$ 2015, under good weather conditions with $\sim$0.8$^{\prime\prime}$ seeing and 10\,mm precipitable water vapor. The spectroscopic calibration was obtained by observing two different Cohen standard stars before and after the science target. The 0.36$^{\prime\prime}$  slit was used with a 6$^{\prime\prime}$ throw.\\
 The IDL pipeline iDealCam \citep{2014ascl.soft11009L}, which was custom-designed to reduce imaging data of Canaricam, was used for the data reduction. 
After removal of the thermal background, the pipeline produces individual frames (or saveset) of $\sim$2\,s duration that can be combined in a long integration sequence. 
Savesets can be realigned along their centroid through two-dimensional Lorentzian\footnote{The FITPSF procedure from F. Varosi's IDL library was used.} fit of the PSF prior to the final shift-and-add stacking. This allows reducing centering and tip-tilt errors that may otherwise lead to unwanted broadening of the PSF. 
\begin{table*}[h]
\centering 
\begin{tabular}[width=\columnwidth]{c r c r r}       
\hline\hline
Object & On source [s]& \% of frame&FWHM$_L$ [$^{\prime\prime}$]  & FWHM$_P$ [$^{\prime\prime}$]\\   
           &                      & selection     &                                                &                                                     \\   \hline
HD\,169414 & 273 (PAH-1) & 69  &0.213$\pm$0.004 & 0.233$\pm$0.003\\
HD\,179218 & 1501 (PAH-1) & 81  &0.232$\pm$0.003 & 0.255$\pm$0.002 \\
HD\,187642 & 314 (PAH-1) & 79 & 0.217$\pm$0.004 & 0.234$\pm$0.003\\
 \hline
HD\,169414 & 287 (PAH-2) & 70  &0.267$\pm$0.002 & 0.282$\pm$0.002\\ 
HD\,179218 & 1651(PAH-2) & 91 &0.280$\pm$0.001 & 0.298$\pm$0.001\\
HD\,187642 & 382 (PAH-2)& 91 &0.268$\pm$0.002 & 0.285$\pm$0.002\\
 \hline
HD\,169414 & 397 (Si-6) & -- & -- & -- \\ 
HD\,179218 &1667 (Si-6) & 90  &0.293$\pm$0.001 & 0.313$\pm$0.002\\
HD\,187642 & 364 (Si-6) & 91  &0.294$\pm$0.002 & 0.311$\pm$0.003\\
\hline \\
\end{tabular}
\caption{Measured mean FWHM and its resulting 3$\sigma $ uncertainty from the distribution of savesets. FWHM$_L$ is estimated through a Lorentzian fit of the saveset's PSF. 
The term $\sigma$ is here the error on the mean of the distribution, that is, the estimated root mean square divided by $\sqrt{N}$ with $N$ being the number of selected frames. 
FWHM$_P$ is estimated graphically from the PSF profile according to the definition of the FWHM. The reported 3$\sigma$ uncertainty is computed from the mean and error on the mean of the distribution of radial profile computed for each saveset. Therefore, the uncertainty obtained with both methods are comparable.}\label{tableFWHM}
\end{table*}
Despite observing in the mid-infrared range with a good $\sim$0.5--0.6$^{\prime\prime}$ average seeing, the atmospheric turbulence above a 10-m class telescope still degrades the image quality, resulting in a non-fully diffraction-limited PSF. Depending on the instantaneous strength of turbulence, the image quality of individual savesets can be significantly affected, which appears in the form of distorted and elongated PSFs. For all our targets, we visually inspected each saveset of the CanariCam dataset and discarded the most corrupted images, that is, where the PSF clearly departs from circular shape. We note that in order to avoid biases by selecting only the ``best'' images, we applied the same visual criterion for frame selection
to both the PSF reference and the science targets. 
The saveset duration of $\sim$2\,s is the same for the science and reference targets, which also have similar fluxes in the N band. 
The resulting on-source integration time for the final stacked images is reported in Table~\ref{tableFWHM} together with the percentage of frame selection. The final stacked image is obtained after re-centering and co-addition of the ``positive'' and ``negative'' images resulting from the data reduction (see Sect.~\ref{profile}). 
Except for the first calibrator in the Si-6 filter for which the seeing conditions were not sufficiently good (see later), the procedure of frame selection resulted in discarding about 20\% of the frames. We inspected approximately 448 savesets for the science target and approximatively 96 savesets for both reference stars and for the three filters. 
The frame selection resulted in an effective on-source total integration time for HD\,178219 target of 
1572\,s, 1653\,s and 1667\,s for, respectively, the PAH-1 PAH-2 and Si-6 filters. 
Figure~\ref{frameselection} illustrates the result of the process of frame selection on the FWHM distributions used in our analysis. 
The FWHM is estimated by fitting of a Lorentzian function. 
Points have been removed when they correspond to a visually distorted PSF. Some points lie below the theoretical diffraction limit of the telescope (dashed line) due to improper Lorentzian fit of the corresponding PSF and have been consequently removed for both the science and PSF calibrators. Importantly, we note that comparatively large values of the FWHM are not removed from the distributions as long as the visual inspection of the corresponding saveset is compatible with a circular PSF. In this way, we aim at limiting bias effects in the analysis of the FWHM distributions.

\section{Observational results}\label{sec:results}
\subsection{Statistics of the full width at half maximum}\label{sec:statistics}
\begin{figure}[t]
\includegraphics[width=\columnwidth]{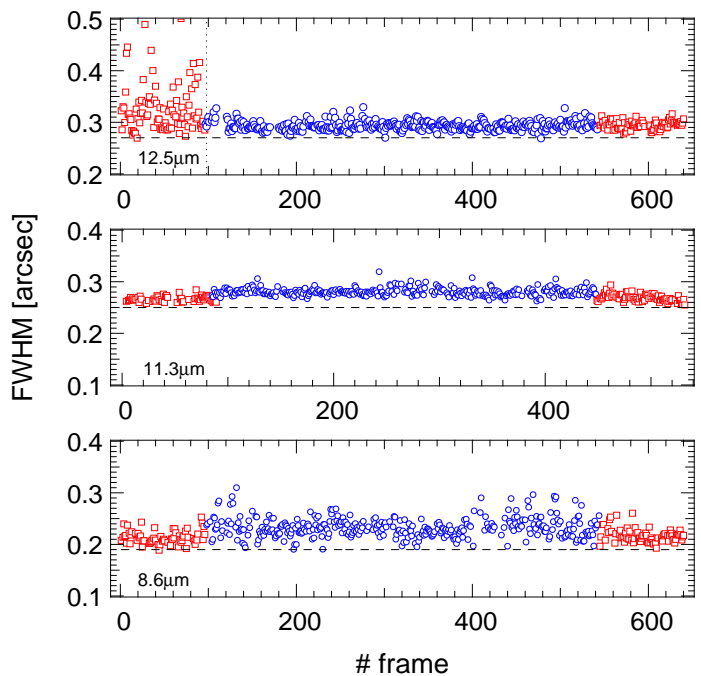}
\caption{Distribution as a function of time of the measured FWHM for the individual savesets. The blue open circles correspond to the science target HD\,179218 and the red open squares correspond to the adjacent calibrators HD\,169414 and HD\,187642. The plots correspond, respectively from bottom to up, to the filters PAH-1, PAH-2 and Si-6. The integration time for the individual saveset for both the science and the calibrator is 2.1\,s (PAH-1), 2.5\,s (PAH-2) and 2.1\,s (Si-6). The horizontal dashed lines show the theoretical diffraction limit of the GTC, which is 0.19$^{\prime\prime}$, 0.25$^{\prime\prime}$ and 0.27$^{\prime\prime}$ at respectively 8.6\,$\mu$m, 11.3\,$\mu$m and 12.5\,$\mu$m, including the central obscuration. 
For the Si-6 filter, the vertical dotted line around frame \#100 shows the effect of poor seeing on the measured FWHM distribution for the first calibrator.}\label{savesetsequence}
\end{figure}

Similarly to \cite{Moerchen2007}, we have explored the statistical behavior of the PSF full width at half maximum (FWHM) distribution obtained after Lorentzian fitting of the individual selected savesets. By treating statistically individual 2s-short images 
and employing sub-pixel recentering for both the science and calibration targets, we minimize the influence of long-term biases (e.g., guiding errors, pupil rotation, and seeing fluctuations) that may result in the broadening of the final PSF when simply stacking the long sequence images. 
We extracted from the distribution of the FWHM data the mean and the error on the mean $\sigma$/$\sqrt{N}$, where $\sigma$ is the distribution standard deviation and $N$ is the number of savesets, respectively nodsets, in the distribution. Figure~\ref{savesetsequence} shows the distribution of FWHM values of the individual savesets for each sequence calibrator--science--calibrator in the three filters. It is already possible to visually discriminate the vertical positioning of the bulk of the distribution for the science (blue open circles) and the adjacent calibrators (red open squares). The plots provide evidence that, for the PAH filters at 8.6\,$\mu$m and 11.3\,$\mu$m, 
the FWHM of HD179218 is on average larger that the FWHM of the adjacent calibrators. 
For the filter Si-6 at 12.5\,$\mu$m, if we neglect the FWHM distribution of the first calibrator corrupted by poor seeing, the FWHM distribution for HD179218 does not exhibit any remarkable deviation from that of the calibrator.\newline
\noindent In Table~\ref{tableFWHM} we report for the saveset distribution the estimated mean FWHM along with the corresponding 3$\sigma$ error for the science and the calibrators in the three different filters. 
For consistency, we compared the FWHM statistics obtained with Lorentzian fitting to the statistics obtained by the direct graphical reading of the FWHM according to its definition. 
We can see that the Lorentzian fit approach systematically gives an average FWHM lower by $\leq$ 10\% than the FWHM derived from a graphical reading. However, the relative trends between science and calibrators are reproducible. 
We observe that within the 99.7\% confidence level the measured FWHM of the science target is larger than the one of the adjacent calibrators for both PAH filters, while it is the same in the Si-6 filter at 12.5\,$\mu$m. 
\noindent We further assessed the resolved nature of HD\,179218's emission by applying the criterion of \cite{Moerchen2010}. Namely we compare the difference in FWHM between the science target and the calibrator star, that is,
\begin{eqnarray}
{\rm FWHM}_{\rm sci}-{\rm FWHM}_{\rm cal}
,\end{eqnarray}
to the combined standard deviation (i.e., error) of the mean
\begin{eqnarray}
\sigma_{\rm tot} =  \sqrt{\sigma^{2}_{\rm cal}+\sigma^{2}_{\rm sci}} 
.\end{eqnarray}

\noindent With ${\rm FWHM}_{\rm sci}-{\rm FWHM}_{\rm cal}\geq 3\sigma_{\rm tot}$ the science source can be considered as spatially resolved. According to this criterion, we give in Table~\ref{fwhm} the result of this analysis and show that the circumstellar emission around HD\,179218 is resolved at $\geq$3\,$\sigma$ confidence level in the two PAH filters, but is unresolved at 12.5\,$\mu$m.

\begin{table}[b]
\centering
\begin{tabular}[\columnwidth]{l c c c}       
\hline\hline
 Filter &  FWHM$_{\rm sci}$\,-\,FWHM$_{\rm cal}$ &  3$\sigma_{\rm tot}$ & Resolved \\ \hline
PAH-1 & 0.019/0.022 (Cal\,1)  & 0.005/0.004 & Y \\
PAH-1 & 0.015/0.021 (Cal\,2)& 0.005/0.004 & Y \\ 
 PAH-2& 0.013/0.016 (Cal\,1)& 0.002/0.002 & Y \\ 
 PAH-2& 0.012/0.013 (Cal\,2)& 0.002/0.002 & Y \\ 
 Si-6 & 0.001/0.002 (Cal\,2)& 0.002/0.004 & N \\  \hline
\end{tabular}
\caption{Resolution criterion for HD\,179218 compared to the adjacent calibrators following \cite{Moerchen2010}. For columns (2) and (3), the first term corresponds to the Lorentzian fit method and the second term corresponds to the direct FWHM graphical read from the PSF profile.}\label{fwhm} 
\end{table}

\subsection{Retrieval of the PSF profile}\label{profile}

The final images are obtained after re-centering and stacking the individual savesets. Image re-centering has to be performed carefully to avoid a significant increase of the final FWHM typically observed in long observing sequences. A simple recentering step based on matching the individual image centroid induces an increase in the FWHM of the final stacked image by $\sim$15\% compared to a single saveset. To improve the centering step, we realigned each saveset by minimizing the quadratic difference between the image of each normalized PSF and the image of the first PSF of the sequence taken as a reference. This operation is implemented with a sampling accuracy of the image of one fifth of a pixel. Such an approach led to a sharper PSF. 
Figure~\ref{Fig:cumulative} gives the Lorentzian-fitted FWHM of the stacked image as a function of the number of coadded savesets. The plots shows, within the limits of the available frames, that the FWHM of the stacked image tends to the statistical value found in Table~\ref{tableFWHM}. \newline
After recentering the individual savesets as described above, the radial profiles for the calibrator and science targets is built for the different available filters. To construct such a profile along with its associated error bars, we extracted the profile for each individual recentered saveset and computed the mean and the error on the mean for each radial pixel. The results for the three different filters are shown in Fig.~\ref{profiles} for the science and calibrator PSF.

\begin{table}[t]
\centering
\begin{tabular}[width=\columnwidth]{ l c c c }       
\hline\hline
Filter &  Flux sci [Jy]  & Cal & Flux cal [Jy]\\ \hline
PAH-1 &  14.9$\pm$0.4 & HD\,169414 & 19.2 \\
PAH-1 &  15.3$\pm$0.4 & HD\,187642 & 43.3 \\ \hline
PAH-2 &  22.4$\pm$0.6 & HD\,169414 & 12.2  \\
PAH-2 &  22.1$\pm$0.4 & HD\,187642 & 25.4\\ \hline
Si-6 &  16.8$\pm$1.7 & HD\,169414 & 9.9 \\
Si-6 &  16.2$\pm$0.7 & HD\,187642 & 20.8 \\ \hline \\
\end{tabular}
\caption{Photometric calibration of HD179218 in the CanariCam filters using the photometric standards reported in column\,(3). The reported errors are 3$\sigma$ uncertainties. The larger uncertainty obtained with HD169414 in the Si-6 filters is due to the initial poorer conditions of the night. We note that the uncertainties reflect only the photometric stability of our measurement. It is however known that many of the Cohen standards show some variability, which results in an absolute photometric accuracy of about 10\%.}\label{photometry}
\end{table}

\subsection{Photometry}\label{sec:photometry}

We performed aperture photometric calibration of HD179218 using HD169414 and HD187642 as Cohen photometric standards. With the two reference stars observed before and after the science target, we can also probe the long term photometric stability of the night. The aperture radius was optimized to 4$\times$FWHM, or $\sim$1$^{\prime\prime}$. 
This size is close to the standard 5$\times$FWHM recommended by photometry manuals. 
The residual background was estimated in a surrounding ring with an inner and outer radius of $\sim$1.6$^{\prime\prime}$ (20 pixels) and $\sim$2.4$^{\prime\prime}$ (30 pixels), respectively. For each science and calibrator target the photometric accuracy is derived from the measurement of the standard deviation of the flux over the savesets time series. The values for the photometric standards in the CanariCam filters are taken from www.astro.ufl.edu/$\sim$dli/web/IDEALCAM$\_$files/iDealCam$\_$v2.0. zip. The results of the photometric calibration are shown in Table~\ref{photometry} and are consistent with Spitzer spectroscopy data by \cite{Fedele2008} and \cite{Juhasz2010}.

\subsection{Spectroscopy}\label{spec}

The spectrum of HD\,179218 was reduced with the RedCan pipeline \citep{Redcan2013} and is shown in Fig. 3. The reduced spectrum is found to overestimate the flux by $\sim$40\% in comparison to our photometric values, which may have different causes (e.g., presence of cirrus or imperfect background subtraction). As we wish to make a relative comparison of the shape of the CanariCam spectrum and the Spitzer spectrum, the former one has been rescaled to the measured photometric values reported in Table 4. 
After rescaling, the relative comparison with the Spitzer spectrum shows good agreement in the shape with the visible peaks at $\sim$8.6-8.7\,$\mu$m, 10.6\,$\mu$m, and $\sim$11.2-11.3\,$\mu$m. Qualitatively, the flux density is slightly larger in the PAH-1 filter for CanariCam than for Spitzer, while it is lower in the PAH-2 filter. 
\begin{figure}[b]
\includegraphics[width=\columnwidth]{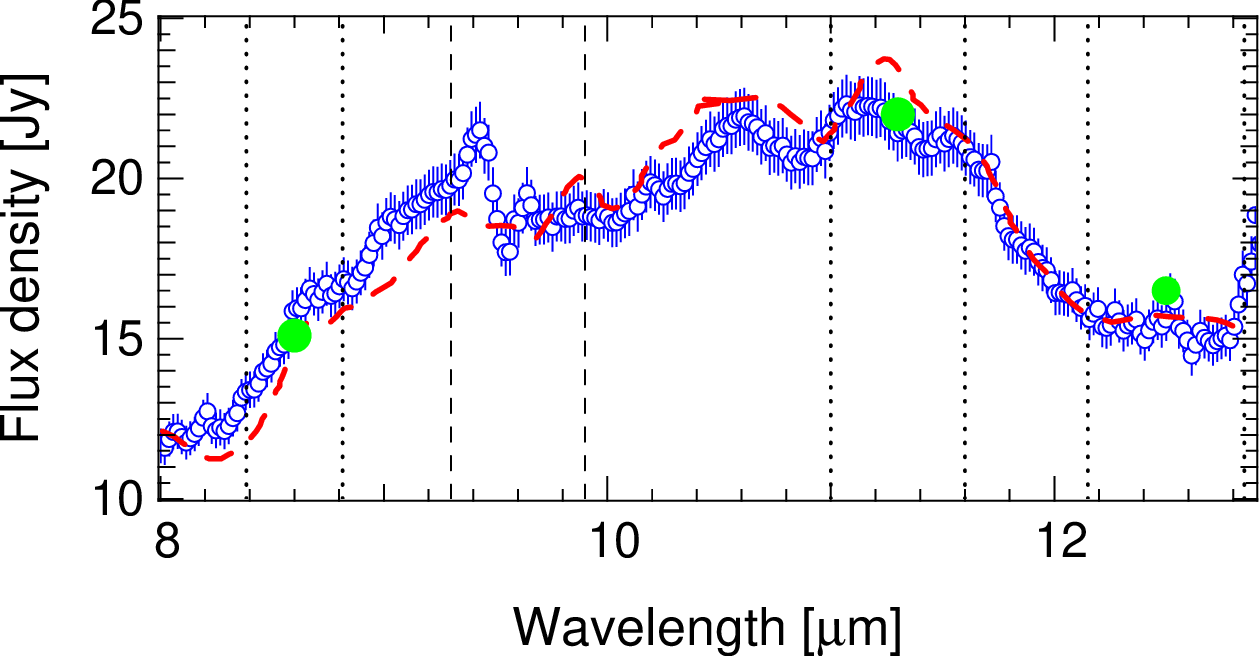}
\caption{Low-resolution spectrum of HD179218 (open circles with error bars) scaled to the measured photometry at 11.3\,$\mu$m (green filled circles). The red dashed line is the corresponding Spitzer/IRS spectrum. The vertical dotted lines denote the bandwidth of the PAH-1, PAH-2, and Si-6 filters. The vertical dashed line indicates the region of the atmospheric ozone feature where proper spectroscopic calibration is cumbersome.
 }\label{spectrum}
\end{figure}
The spectral calibration in the 9.3--9.9\,$\mu$m region is strongly affected by the ozone atmospheric feature and residuals of the data reduction can be seen. We also remark that the flux measured by CanariCam between 9 and 9.2\,$\mu$m appears overestimated by 10 to 20\% in comparison to ISO and Spitzer.

\section{Characteristic size of the emission}\label{characteristicsize}
\subsection{Quadratic subtraction of FWHMs}\label{quadsub}

For small differences in FWHM measurements, as in our case, PSF deconvolution is a delicate technique which strongly depends on assumptions made on the PSF and number of iterations. 
Similarly to \cite{Moerchen2010, Marinas2011}, an alternative to deconvolution is the estimate of the disk diameter from the quadratic subtraction of the science PSF and calibrator PSF derived in Sect.~\ref{sec:results} through 
$D^2_d$ = FWHM$^2_{sci}-$ FWHM$^2_{cal}$, where $D_d$ is the estimated disk diameter. 
The errors associated to $D_d$ are calculated following Eq.~2 in \cite{Marinas2011}. 
Our estimates are reported in Table~\ref{sizes}. 
Assuming a distance of 293\,pc (cf. Sect.~\ref{modeling}), we found in the two filters a comparable characteristic diameter of $\sim$24--30\,au. On average, the disk emission appears slightly more extended in the PAH-1 filter than in the PAH-2 filter. The disk emission is found to be unresolved in the Si-6 filter in the sense of the 3$\sigma_d$ criterion.

\subsection{Gaussian disk}

\begin{table}[t]
\centering
\begin{tabular}[width=\columnwidth]{l c c c}       
\hline\hline
 Calibrator &  Filter &  $D_{d,L}$\,[$^{\prime\prime}$]  & $D_{d,PSF}$\,[$^{\prime\prime}$]  \\ \hline
HD\,169414 & PAH-1 &  0.092$\pm$0.004 & 0.103$\pm$0.003\\ 
HD\,187642 & PAH-1 &  0.082$\pm$0.005 & 0.101$\pm$0.003 \\ 
HD\,169414 & PAH-2 &  0.084$\pm$0.003 & 0.096$\pm$0.002 \\ 
HD\,187642 & PAH-2 &  0.081$\pm$0.002 & 0.087$\pm$0.002 \\ 
HD\,187642 & Si-6 &  $\leq$0.024$\pm$0.009 & $\leq$0.035$\pm$0.011 \\ \hline \\ 
\end{tabular}
\caption{Angular diameter of the circumstellar emission measured in the three different filters. The uncertainty refers to the 1$\sigma_d$ error as derived in \cite{Marinas2011}. 
The source is considered resolved if the deconvolved (in the sense of quadratic subtraction) diameter is larger than 3$\sigma_d$. 
The subscript $L$ and $P$ refer to the Lorentzian fit and to a direct measurement of the PSF profile, respectively.}\label{sizes} 
\end{table}

We complemented the previous estimate with a simple approach to determine the characteristic size of the resolved emission. Namely, we model the emission as a two-dimensional face-on Gaussian disk convolved with the telescope PSF in the corresponding filters. This model only depends on the FWHM of the Gaussian function, though it might not be always sophisticated enough to reproduce the full shape of the PSF profile (core+wings). 
The characteristic size is estimated by visually matching the synthetic profile to the science profile within the experimental error bars. In this analysis we used the final images after recentering and stacking. \\
In the PAH-1 filter, a Gaussian disk with a FWHM of 95$\pm$6\,mas reproduces our science PSF profile, whereas a Gaussian disk with a FWHM\,$\leq$22$\pm$7\,mas would remain spatially unresolved. In the PAH-2 filter, the Gaussian disk model fitting our science profile has a FWHM of 101$\pm$7\,mas, and the unresolved disk would have a FHWM\,$\leq$25$\pm$7\,mas. Finally, in the Si-6 filter the Gaussian disk must have a FWHM\,$\leq$36$\pm$7\,mas 
so as not to exceed the FWHM of the science profile. This analysis confirms that the disk emission in HD179218 is resolved in the PAH filters and unresolved in the 12.5$\mu$m Si-6 filter. However, it is not possible to conclude within the error bars on a difference in the angular size of the emission in the PAH-1 and PAH-2 filters.

\section{Modeling}\label{modeling}

Our imaging results show that the circumstellar emission around HD\,179218 is spatially resolved in the two PAH filters whereas it remains unresolved at 12.5 $\mu$m. We attempt to identify the origin of the extended emission. A natural comparison arises with the case of HD96048 for which \cite{Lagage2006} resolve the emission of polycyclic aromatic hydrocarbons at the surface of the flared disk in direct view of the central star. 
In order to test this possible configuration in the case of HD\,179218, we adopt the following strategy: we develop a radiative transfer model of a disk that simultaneously fits the SED and is spatially unresolved 
in the synthetic CanariCam image at 12.5\,$\mu$m. 
The idea is to constrain the disk's size in the 12.5\,$\mu$m band, which is dominated by the dust thermal emission and shows no significant presence of PAH emission \citep{Acke2010, Juhasz2010}. 
A PAH-free disk emission model with the same parameters is then produced at 8.6 and 11.3\,$\mu$m from which a synthetic observational profile can be extracted and compared to our observations. 
A similar strategy was successfully used by \cite{Honda2012} to model the gap's size of HD169142 in the Q band. 
In this approach, the modeling of the thermal emission at 12.5\,$\mu$m gives us an upper limit on the outer disk's dimension while the inner regions are unresolved with CanariCam. If the 8.6 and 11.3\,$\mu$m synthetic profiles are spatially unresolved, this would be a good indication that the observed resolved emission is not of thermal origin (in the sense of ``continuum'' origin). 
On the contrary, in case the 8.6 and 11.3\,$\mu$m synthetic profiles are found to be spatially resolved, further observational constraint needs to be added through, for instance, mid-infrared interferometric data to be conclusive on the nature of the resolved emission in the PAH bands.\\
\cite{Dominik2003} proposed a first radiative transfer model based on a single disk geometry with an outer radius of 30\,au and a positive power-law index $p$=2 of the surface density. 
In a statistical study of HAeBe's disks, \cite{Menu2015} used simple geometrical temperature gradient models in combination with mid-infrared interferometry data to infer a half-light radius of the disk of 7$\pm$1.2\,au at 254\,pc. 
Using nulling interferometry, \cite{Liu2007} derived a radius of 13.5$\pm$3\,au based on a ring-like disk model at a similar distance of 244\,pc. Moreover, \cite{Fedele2008} applied an achromatic geometrical disk model to VLTI/ MIDI data and proposed as their best solution a two-component pre-transitional disk structure with an inner disk extending from 0.3--3\,au and an outer component whose  bulk mid-IR emission lies in a 13--22\,au region at 240\,pc.

\subsection{Description of the disk model}\label{zdiskmodel}

\begin{table}[t]
\centering
\begin{tabular}[width=\columnwidth]{c c c c c c }       
\hline\hline
$L_{\ast}$\,[$L_{\odot}$] & $M_{\ast}$\,[$M_{\odot}$] & $T_{\ast}$\,[$K$] & $R_{\ast}$\,[$R_{\odot}$] & A$_V$ [mag] & $d$\,[pc]  \\ \hline
180$^{(a)}$ & 3.66$^{(a)}$ & 9640$^{(a)}$ & 4.8$^{(a)}$ &0.63$^{(b)}$ & 293$^{(c)}$ \\ \hline
\end{tabular}
\caption{Stellar parameters assumed in this study. ({\it a}): extracted from \cite{Alecian2013}; ({\it b}):  fit of the optical photometry; ({\it c}): GAIA parallax.}\label{baseline} 
\end{table}

We developed radiative transfer disk models for HD\,179218 aiming at constraining simultaneously the spectral energy distribution (SED) and the imaging data on the source. We used for this purpose the well-established Monte-Carlo code RADMC3D \citep{Dullemond2004} that permits one to synthesize disk images and SEDs. 
Assuming a disk in vertical hydrostatic equilibrium and perfect gas/dust coupling, the dust density in g.cm$^{-3}$ is modeled analytically according to

\begin{eqnarray}
\rho(r,z)&=&\frac{\Sigma(r)}{H(r)\sqrt{2\pi}}\exp\left [ -\frac{1}{2}  \left ( \frac{z}{H(r)} \right )^2   \right ]\label{Eq3}
,\end{eqnarray}

\noindent where $\rho(r,z)$ is described via the parametrized dust surface density $\Sigma(r)$=$\Sigma_{\rm out}(r/r_{\rm out})^p$ and dust scale height $H(r)$=$H_{\rm out}(r/r_{\rm out})^{(1+\beta)}$, and $p$ is the surface density exponent and $\beta$ the disk flaring exponent. The subscript {\it out} refers to the outer radius of each disk component considered.\\
A difference with respect to earlier models is that we assume more recent estimates of the stellar parameters and parallax, which naturally influences the radiative transfer calculation and the production of the synthetic images. 
We used for the central star the parameters from \cite{Alecian2013}, namely a luminosity $L_{\ast}$=180$L_{\odot}$, a radius $R_{\ast}$=4.8$R_{\odot}$, a mass $M_{\ast}$=3.66$M_{\odot}$ and an effective temperature $T_{\rm eff}$=9640\,K. A luminosity of 80--100$L_{\odot}$ was typically assumed in the earlier works. 
Based on a recent GAIA parallax measurement\footnote{http://gea.esac.esa.int/archive/} of 3.41$\pm$0.35\,mas, we assume a distance of 293\,pc rather than the 240\,pc found in the literature. 
The experimental SED of the system is taken from \cite{Acke2004} and contains photometric data from the literature as well as spectroscopic data from ISO. \\
The RT grid extends from the dust sublimation radius to 200 AU. The disk temperature distribution is computed through a first Monte Carlo run using 10$^6$ photon packets. Since isotropic scattering is considered in our modeling, the scattering source function is then computed at each wavelength through an additional Monte Carlo run using 3$\times$10$^5$ (resp. 5$\times$10$^4$) photon packets for the images (resp. for the SED). 
A ray tracing method is then applied to compute the synthetic SED and images (at 8.6, 11.3, and 12.5\,$\mu$m). In our RT modeling, we do not include PAHs, although they are clearly present, and model only the dust continuum. This is discussed later in the paper. 
\newline
Following \cite{Fedele2008}, we assume as a starting baseline a passive irradiated pre-transitional disk structure consisting of an inner narrow ring, a low dust density gap region and a larger outer disk. The outer component is decomposed into a warm disk atmosphere component that will dominate the mid-IR emission, and a colder mid-plane component optically thick at 10\,$\mu$m that will dominate the far-IR and sub-mm emission. 
The outer disk warm atmosphere will mostly influence our data in the mid-infrared (see Sects.~\ref{sec53} and \ref{sec54}).

\begin{table*}[h]
\centering
\begin{tabular}[width=\pagewidth]{l  c c c c c c c c }       
\hline\hline
                 &  $r_{\rm in}$\,[au] & $r_{\rm out}$\,[au] & $H_{\rm out}/r_{\rm out}$ & $M_{\rm dust}$\,[$M_{\odot}$]$^{(\it c)}$& $p$ & $\beta$ & $i$\,[$^{\circ}$]$^{(\it d)}$ & P.A.\,[$^{\circ}$]$^{(\it d)}$  \\ \hline
Inner Disk &1.1$^{(\it a)}$ & 3$^{(\it a)}$ & 0.02$^{(\it a)}$ &  1.46$\times$10$^{-5}$$^{(\it a)}$  & 0$^{(\it a)}$ &1/7 & 57 & 23\\ 
Gap & 3.1$^{(\it a)}$ & 9.9$^{(\it a)}$ & 0.1$^{(\it a)}$ & 1.83$\times$10$^{-13}$$^{(\it a)}$&  0$^{(\it a)}$  &  1/7 & 57 & 23  \\ 
Outer Disk &  10$^{(\it a+b)}$ & 80$^{(\it a+b)}$ & 0.12$^{(\it a)}$ & 1.46$\times$10$^{-4}$$^{(\it a)}$ & -1.5$^{(\it a+b)}$ & 1/7 & 57 & 23 \\ \hline
\end{tabular}
\caption{Best-model parameters resulting from the $\chi^2$ minimization. Superscript (a) denotes the parameters fitted from the SED. Superscript (b) denotes the parameters fitted from the 12.5\,$\mu$m PSF profile. The parameters $\beta$, $i$ and P.A. are not fitted. ({\it c}): the surface density $\Sigma_{\rm out}$ in Eq.~\ref{Eq3} is parametrized through the mass $M_{\rm dust}$. A gas-to-dust ratio of 100 is assumed. ({\it d}): from \cite{Fedele2008}.
}\label{model} 
\end{table*}

\subsection{The dust opacities}

The continuum emission being entirely dominated by the dust, the grain composition and resulting opacity influences the shape and values of the SED from the near-IR to the sub-mm. As we do not aim at a detailed fit of the spectral feature already done elsewhere, our approach is to assume opacity laws detailed in the literature in order to place ourselves in a realistic case. For the outer disk, we assumed a composition of 90\% amorphous silicate grains and 10\% crystalline enstatite grains following the findings of \cite{Juhasz2010}. We assumed a size distribution $\propto$\,$a^{-3.5}$ \citep{Mathis1977} from 0.1\,$\mu$m to 100\,$\mu$m for the amorphous grain population and a fixed size of 2\,$\mu$m for the enstatite grain population. The midplane is populated with larger amorphous silicate grains to reflect dust sedimentation, with sizes ranging from 10\,$\mu$m to 1\,mm following a similar power-law size distribution to the one before. 
Finally, the inner disk/gap is populated with a mixture of amorphous silicate and highly refractory carbon grains at a ratio of approximately 9:1, respectively, in agreement with \cite{Dominik2003}. 

\subsection{Fitting procedure: SED and 12.5\,$\mu$m image}\label{sec53}

The procedure consists in best-fitting the SED through a $\chi^2$ minimization and verifying a posteriori that the corresponding synthetic image is unresolved at 12.5\,$\mu$m. 
For a given radiative transfer model, we thus produced a synthetic SED to be compared to the observational SED, and a synthetic disk image at 12.5\,$\mu$m that we convolved with our PSF reference star in the Si-6 filter. A radial profile is then extracted to be compared with the observed profile at 12.5\,$\mu$m. \newline
As a first step, we run our radiative transfer code to identify a reference baseline model (RBM) that provides a good visual fit to the SED without PAH. This model assumes the properties aforementioned (central star, mineralogy) as well as a pre-transitional disk structure (inner disk + gap + outer disk) \citep{Fedele2008}. A first exploration of the parameters shown in Table~\ref{model} (cf. caption) allows us to converge towards a possible solution for the RBM based on the SED fit. 
As a second step, on the basis of this RBM, we determined which parameters influence most significantly the mid-IR profiles and refine our search on these parameters by including the information on the 12.5\,$\mu$m PSF profile. In this way, we avoid varying all the parameters of the model to minimize degeneracy effects.\\
a) Inner disk: the inner radius is fixed at 1.1AU, which roughly fits the dust sublimation radius given our stellar parameters. We then tested the influence of the outer radius of the inner disk by varying it out to 5\,au and varying the exponent of the power law from $p$=-2 to $p$=2. This corresponds to a range of exponents typically found for disk models \citep{Dominik2003}. We observe a mild influence of these parameters on the SED in the 2--3\,$\mu$m near-IR region and no impact on the PSF profiles at 8.6, 11.3, and in particular at 12.5\,$\mu$m. The inner disk scale height and flaring index have also no measurable impact on the mid-infrared profiles. 
\\
b) Gap: this low-density region hosts a dust mass of $\sim$10$^{-13}$$M_\odot$ in the RBM. The power-law exponent of the surface density was varied from -2 to +2. This parameter did not impact the various PSF profiles either.\\
c) Outer disk: 
the inner and outer radii of the outer disk component, $R_{\rm i}$ and $R_{\rm o}$, along with the surface density power-law exponent were found to influence most significantly the PSF profiles at 12.5 um and the SED at mid-IR up to sub-mm wavelengths. We therefore concentrated on these three parameters in what follows.

\subsection{Exploration of the outer disk's parameters $p$, $R_{\rm i}$ , and $R_{\rm o}$}\label{sec54}

\begin{figure}[t]
\includegraphics[width=\columnwidth]{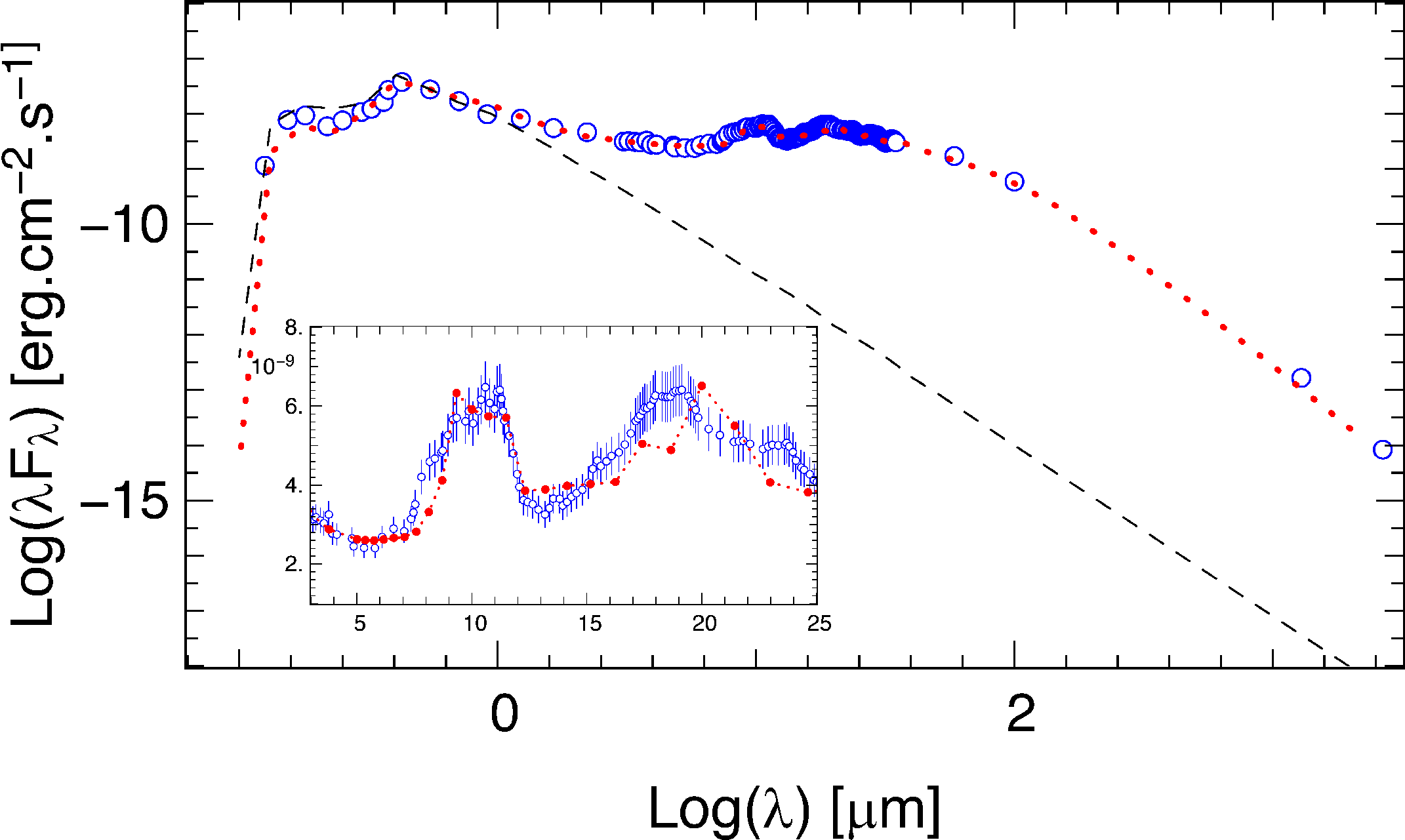}
\caption{Spectral energy distribution of HD\,178219 (blue open circles) and overplotted synthetic SED from radiative transfer modeling (red dotted line) with a resulting $\chi^2_r$=1.9. The line-of-sight extinction is $A_v$=0.63 and the distance $d$=293\,pc. The photometry error bars are conservatively assumed to be 10\%. The inset shows a view on the 3-25\,$\mu$m region of the SED. The black dashed line is the stellar photosphere.
}\label{mysed}
\end{figure}

We have conducted a small parameter search by varying the power law $p$ in the range \{-2,+2\} in steps of 0.5, the inner radius of the outer disk $R_{\rm i}$ in the range \{8\,au,12\,au\} and the outer radius of the outer disk $R_{\rm o}$ in the range \{30\,au,150\,au\}. 
We have simultaneously compared our synthetic SED and 12.5-$\mu$m PSF profile to our observations. Table~\ref{tab:chi2} gives the value of the non-reduced $\chi^2$ for the SED fit as a function of ($p$, $R_{\rm i}$, $R_{\rm o}$) and highlights the models for which the PSF profile at 12.5\,$\mu$m is either spatially resolved or unresolved. The parameter $p$ in any model has the same value for the two components of the outer disk, that is, the disk surface and the midplane.\\
For $p\geq$0, that is, when most of the mass is located in the outer regions of the disk, the PSF profile at 12.5$\mu$m is systematically resolved (red-box values in Table~\ref{tab:chi2}). For a negative power law, we find that an overly small outer radius $R_{\rm o}$ does not produce a satisfactory fit of the SED. For large values of $R_{\rm o}$ (e.g., 150\,au) and a negative power law, the flux in the far-IR and sub-mm range tends to be underestimated. 
A best fit for the SED is found for $p$=-1.5, $R_{\rm o}$=80\,au, and $R_{\rm i}$=10\,au with a non-reduced $\chi^2$ value of 302 (or $\chi^2_r$=1.9 for $N$-$\nu$=157). This model, for which the 12.5\,$\mu$m PSF is unresolved, is then chosen as the reference baseline model for characterizing the PSF profiles in the PAH filters.\\
We highlight that the approach adopted to isolate a disk model for further analysis holds some limitation: For the outer disk, other parameters may influence the result, such as, for example, the flaring index $\beta$ not included in the minimization process. 
Nevertheless, during the search for a reference baseline model, we examined the influence of the flaring index and found that as soon as $\beta$ reaches 2/7, the infrared excess at 5\,$\mu$m is overestimated. 
\begin{figure*}
\includegraphics[width=0.7\columnwidth]{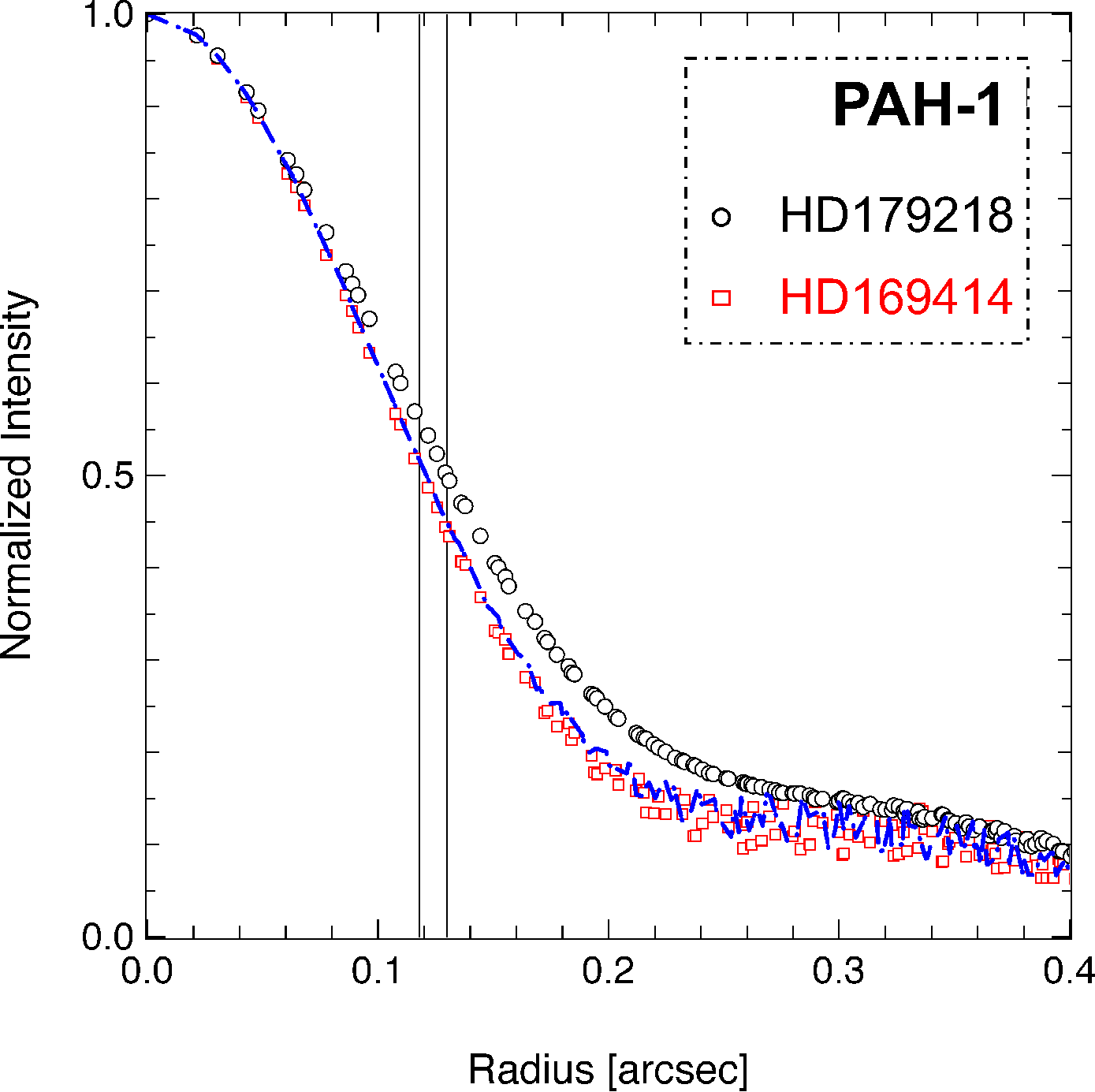}
\includegraphics[width=0.7\columnwidth]{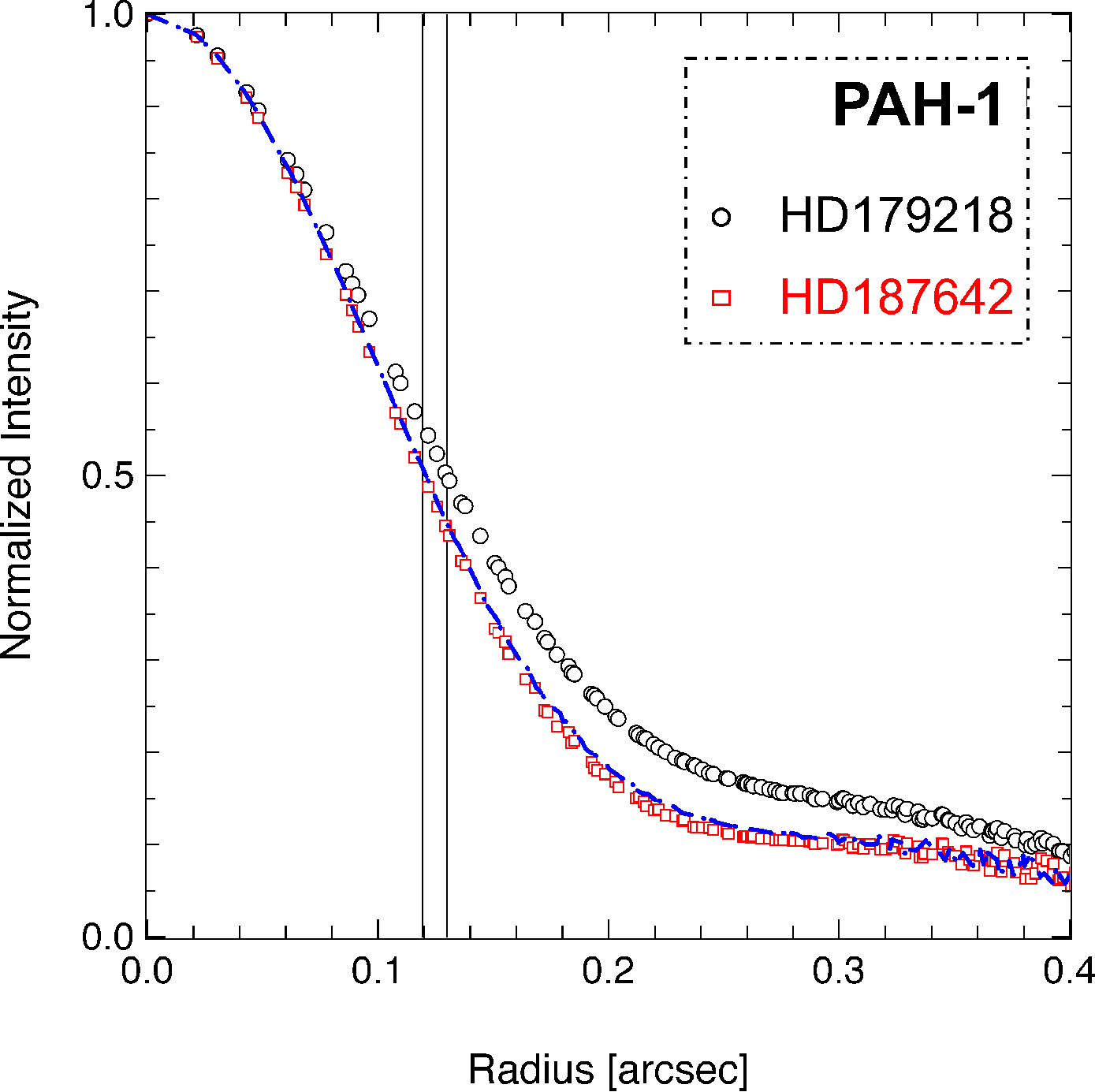} \\
\includegraphics[width=0.7\columnwidth]{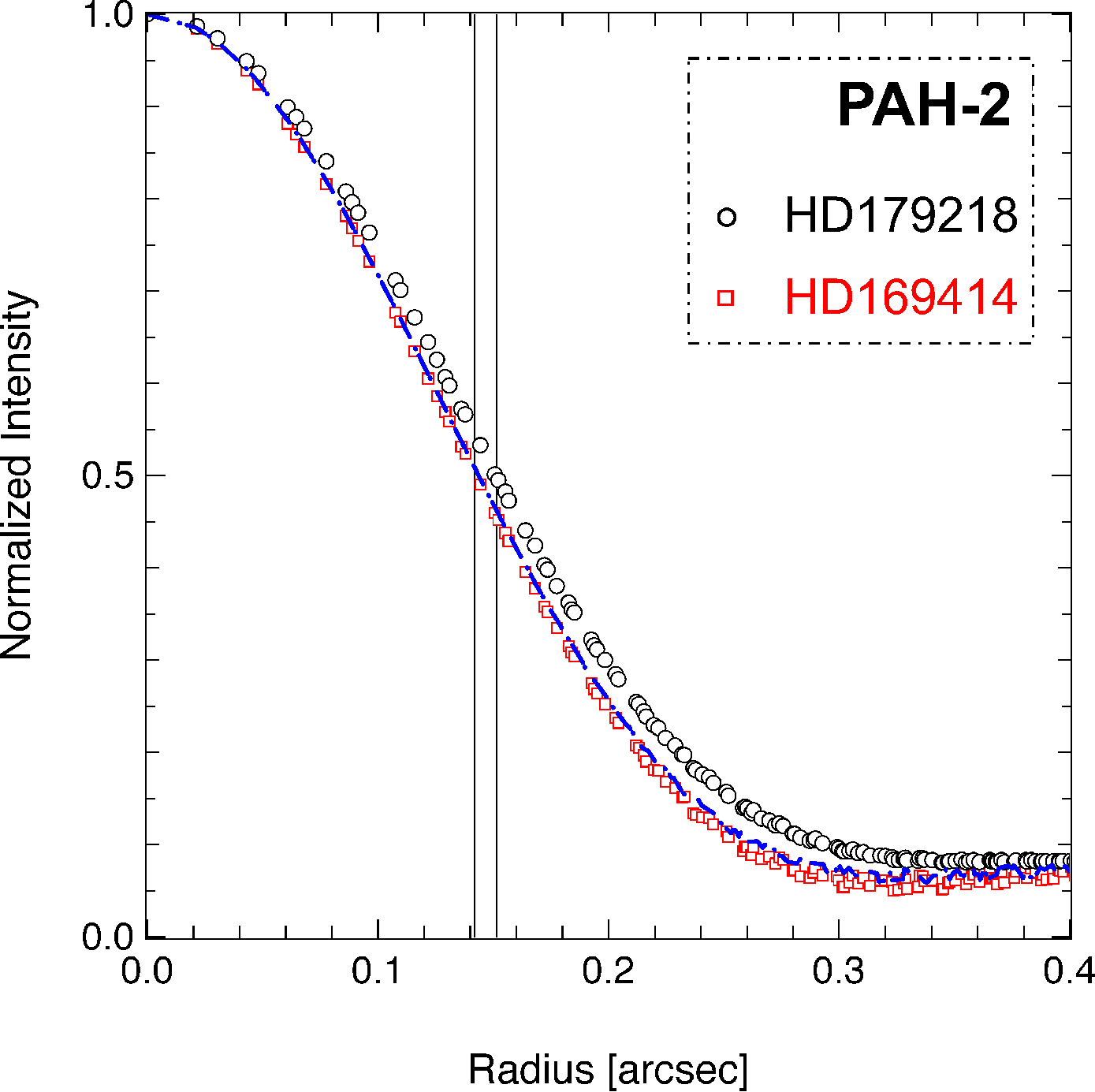} 
\includegraphics[width=0.7\columnwidth]{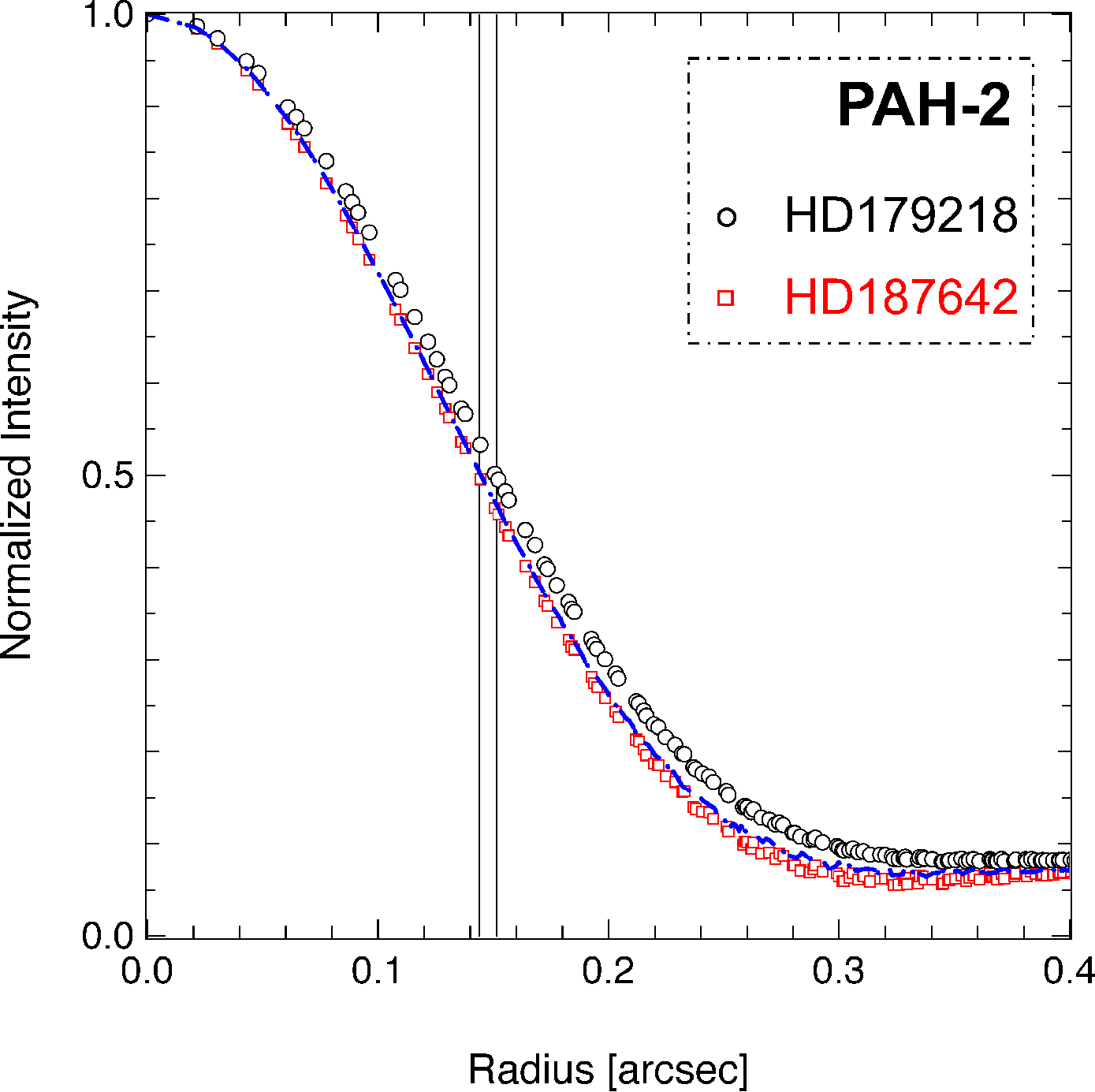}
\includegraphics[width=0.7\columnwidth]{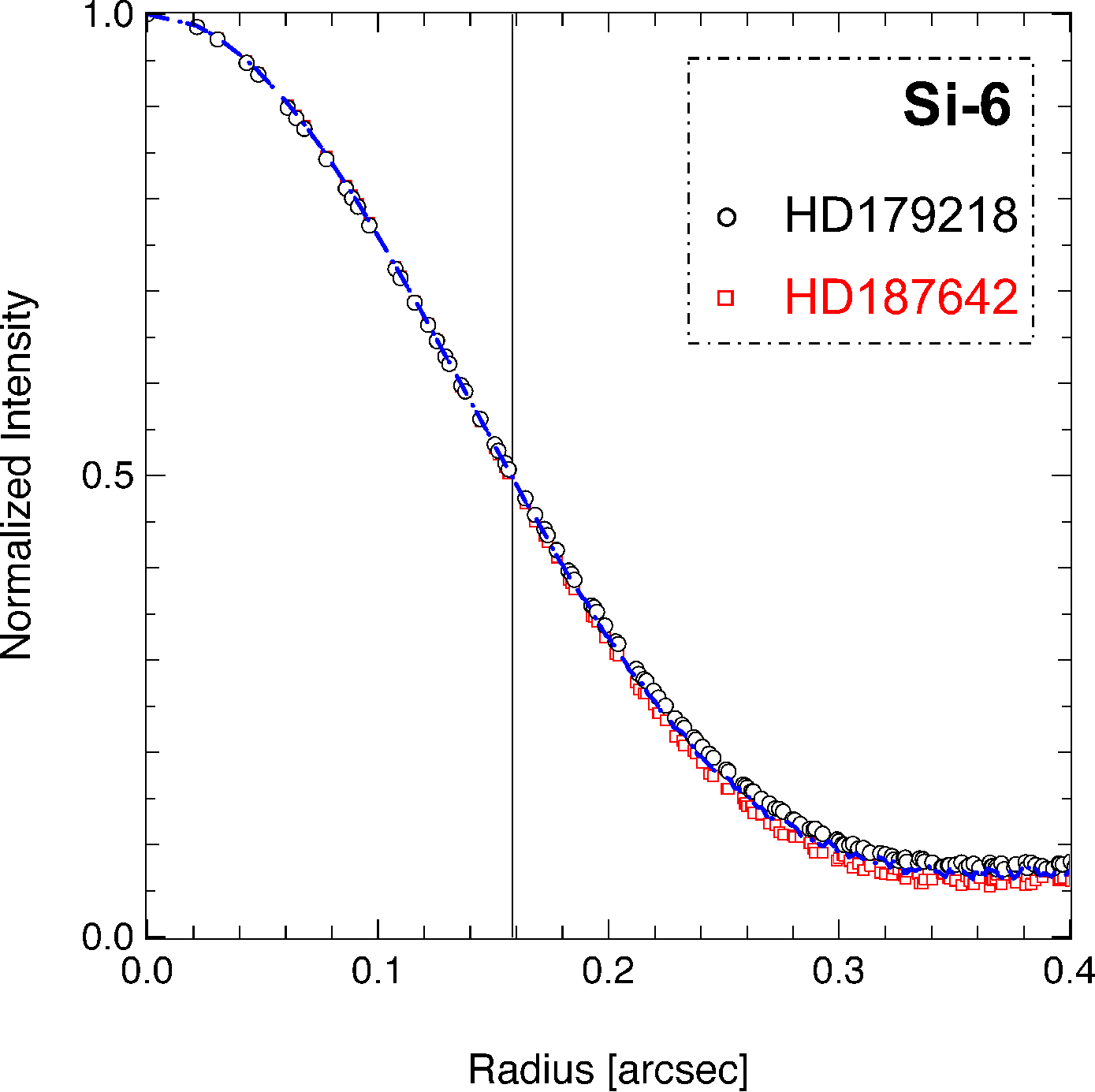}
\caption{
Overview of the science, calibrator, and model PSF profiles. 
In all views, the open red squares refer to the calibrator sources  and the open black circles correspond to the science target. 
For each filter, PAH-1 and PHA-2, the left plot corresponds to the comparison between the science and the first calibrator PSF profiles, whereas the right plot compares the science and the second calibrator PSF profiles. The size of the error bar for each point is similar to the size of the corresponding symbol. 
The dashed blue line is the radial profile obtained after convolution of the corresponding PSF calibrator with the radiative transfer model derived in Sect.~\ref{modeling}. The vertical continuous lines show the difference in FWHM between the calibrator and the science for the values reported in Table~\ref{tableFWHM}.}\label{profiles}
\end{figure*}
This effect cannot be satisfactorily compensated by a reduction of the disk's mass, which would then result in an emission deficit in the mid-IR and sub-mm ranges. The value of $\beta$=1/7 remains conservative in comparison to values in the literature. Figure~\ref{mysed} presents the resulting SED overplotted with the observational one, which show very good agreement over the whole spectral range.

\subsection{Analysis of the PSF profiles}

The best-model parametrized in Table~\ref{model} is used to produce synthetic images at 8.6\,$\mu$m, 11.3\,$\mu$m, and 12.5\,$\mu$m that are then convolved with the GTC PSF, and for which profiles are compared to our observations. Figure~\ref{profiles} presents the result of the comparison in the three different filters. 
The size of each symbol corresponds to roughly the 3-$\sigma$ error on the mean of each radial point of the PSF.
\\
We observe experimentally that the observed science PSF profile (black empty circles) is spatially resolved with respect to the calibrator PSF (red empty squares) between $\sim$0.1 and 0.3$^{\prime\prime}$. This is also clearly repeatable when using the two nearby calibrator stars (see caption). This is observed in the two PAH filters and the continuous vertical lines correspond to the FWHM values established in Table~\ref{tableFWHM} for the science and calibration targets. On the contrary, the observed profiles in the Si-6 filter at 12.5\,$\mu$m do not show a detectable difference in their FWHM, suggesting that the disk's extended emission is not resolved by CanariCam in the thermal continuum. We note that this comparison for the Si-6 filter is made only with the PSF calibrator second in time, as the first one suffered from poorer observing conditions. We also observe that the difference in FWHM between the science and the calibrator PSFs is more prominent in the PAH-1 filter than in the PAH-2 filter.\\
In order to better understand the spatial properties of the thermal emission in different bands, we compare the synthetic PSF profiles derived from the best-fit model of Table~\ref{model} with our observations (blue dash-line). We clearly see that this model of thermal emission has the same profile as the PSF calibrator and is likely unresolved in our three observing bands. It is possible that another source of emission needs to be invoked to explain our observations.

\section{Discussion}

\subsection{Spatially resolved PAH circumstellar emission}

Our measurements show spatially resolved emission in the PAH-1 and PAH-2 filters centered on the infrared emission bands (IEBs) at 8.6 and 11.3\,$\mu$m. These correspond to the two most prominent PAH bands in the 8-13\,$\mu$m region of the spectrum of HD\,179218. The bulk of the emission detectable by our observations has a spatial extent of $\sim$12 to 15\,au  in radius, assuming $d$=293\,pc. No disk emission is resolved in the 12.5\,$\mu$m filter, where the emission is dominated by the dust thermal continuum according to the spectrum \citep{Juhasz2010}. 
Advantageously, the angular diameter of the derived Gaussian disk model of HD\,179218 can be compared to other existing interferometric measurements using the same model. Our Gaussian FWHM of 95$\pm$6\,mas and 101$\pm$7\,mas in the PAH-1 and PAH-2 filters, respectively, is larger than the Gaussian FWHM of 80$\pm$3\,mas at 10.7\,$\mu$m ($\Delta\lambda$=1.45\,$\mu$m) measured for the disk's thermal continuum by \cite{Monnier2009} using aperture masking. Nulling interferometry measurements by \cite{Liu2007} revealed a Gaussian FWHM of 81$\pm$16\,mas at 10.6\,$\mu$m, although over a wide 50\% bandpass encompassing the whole N band. 
Finally, our upper limit measured in the Si-6 filter appears very coherent with the Gaussian FWHM of 0.034$^{\prime\prime}$$\pm$3\% estimated by \cite{Leinert2004} at 12.5\,$\mu$m using MIDI/VLTI. 
Previous single-aperture mid-infrared imaging observations did not resolve at 11.6\,$\mu$m ($\Delta\lambda$\,=\,1.1\,$\mu$m, \cite{Marinas2011}) and in the Q\,band \citep{Marinas2011,Honda2015}. These observations were however conducted with Gemini and Subaru, which deliver intrinsically poorer spatial resolution than the GTC by a factor $\sim$1.2 and were not done in the PAH filters.\\
In a second step, our radiative transfer modeling helped to investigate the nature of the resolved emission. Fitting both the SED data and our imaging data with a model containing a passive disk suggests that the dust thermal emission alone would not be resolved by our CanariCam observations. We suggest that the detected resolved emission is caused by PAH molecules UV-excited by the central star and located near the surface of the flared disk. 
An interesting comparison can be advanced here with HD\,97048, a notable Herbig star for which the PAH emission at 8.6 and 11.3\,$\mu$m is resolved out to several tens of astronomical units \citep{Lagage2006}. 
\cite{Doucet2007} estimated the disk diameter $D_d$ of HD\,97048 - in the sense of the quadratic subtraction used above - to be $\sim$40\,au at a distance of 180 pc, 
which is slightly larger than for HD\,179218. 
However, HD\,97048 being about two times closer, the disk is better resolved in its outer regions. Despite a relatively narrow science PSF core (about 1.2 times larger than the PSF calibrator core), a large amount of PAH emission is found in the wings of HD\,97048's PSF out to 380\,au in radius when comparing it with the emission in the immediate nearby continuum (SIV filter at 10.49\,$\mu$m, see Fig. 3 in \cite{Doucet2007}).
\begin{figure}[t]
\includegraphics[width=0.49\columnwidth]{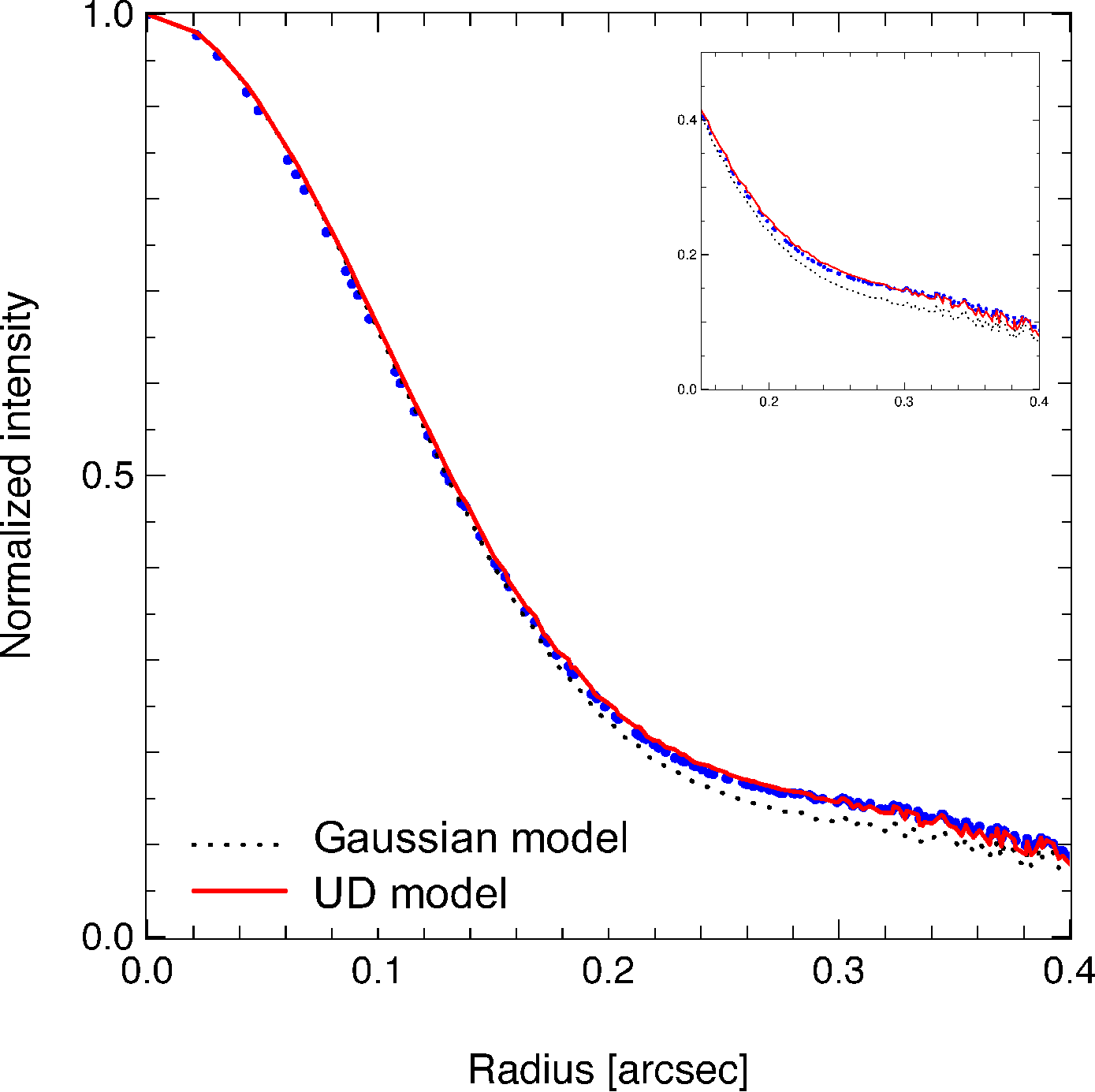}
\includegraphics[width=0.49\columnwidth]{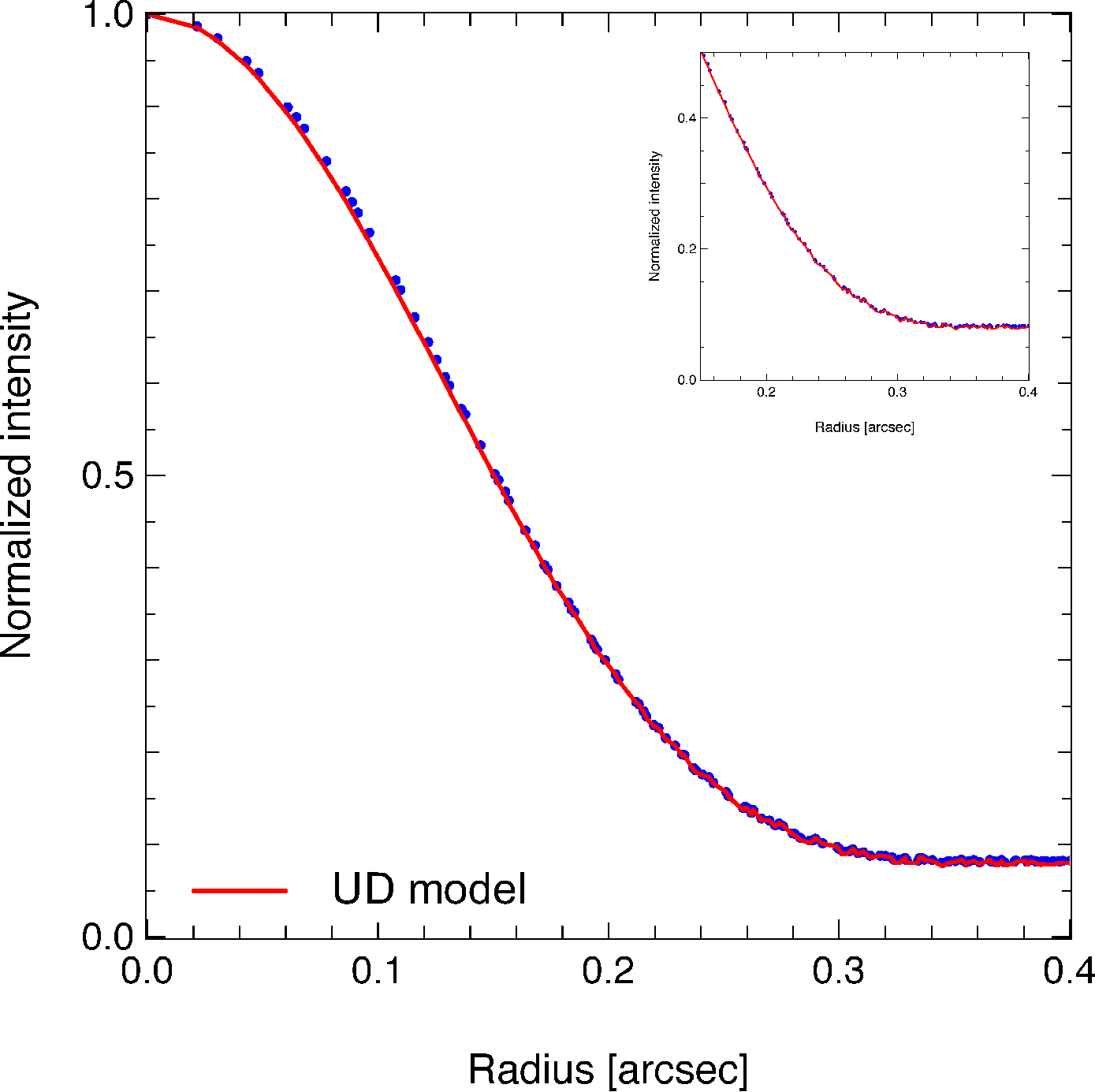}
\caption{Modeling of the science PSF profile 
(blue filled large dots) in the PAH-1 (left) and PAH-2 (right) filter with our hybrid model based on a Gaussian disk (black dotted line) and on the uniform disk (continuous red line) models, respectively. The insets show a close view on the wings of the PSF.}\label{scaledmodel}
\end{figure}\\
Similarly, in the case of HD\,179218, the analysis presented in Sect.~\ref{characteristicsize} may only provide the characteristic size of the disk's emission, but does not allow to constrain the true physical spatial extent of the PAH emission with respect to the continuum emission. For this purpose, it is necessary to account for the flux ratio between the PAH and continuum components in each filter. This can be estimated by subtracting from our photometric measurements in Table~\ref{photometry} the flux density of the PAH contribution at 8.6 and 11.3\,$\mu$m as estimated by Juh\'asz et al. (2010, Table~9). We derive a continuum flux density of 11.4\,Jy and 19.1\,Jy for a corresponding PAH flux density of 
3.9\,Jy 
and 2.9\,Jy, respectively at 8.6 and 11.3\,$\mu$m. We then build a hybrid model based on the image of the HD\,179218's system in the continuum obtained by radiative transfer simulations (cf. Sect.~\ref{modeling}) to which we superimpose a geometrical model simulating the PAH brightness distribution. The relative flux density of each component is scaled accordingly. For the PAH emission component, we investigated a Gaussian model and uniform disk (UD) modified model with radius R$_{UD}$ modulated by a radial power law $r^p$. Using the former model, we are able to fit the core of the science PSF but we underestimate the emission in its wings in particular at 8.6\,$\mu$m, whereas the smoother radial profile of the latter UD model better reproduces the full science PSF profile: As presented in Fig.~\ref{scaledmodel}, we find that in both filters a classical UD model with $p$=0, R$_{UD}$=0.3$^{\prime\prime}$ (87\,au at 293\,pc) and scaled to the corresponding PAHs flux density successfully matches the science PSF profiles of HD\,179218. This result shows that, despite the small characteristic size -- in the sense of quadratic subtraction -- of HD\,179218's disk emission, the PAHs emission at 8.6 and 11.3$\mu$m extends comparatively out to larger radii, which suggests a spatial scenario of the disk's emission similar to HD\,97048.

\subsection{Charge state of PAHs in HD\,179218}
\begin{figure}[t]
\includegraphics[width=\columnwidth]{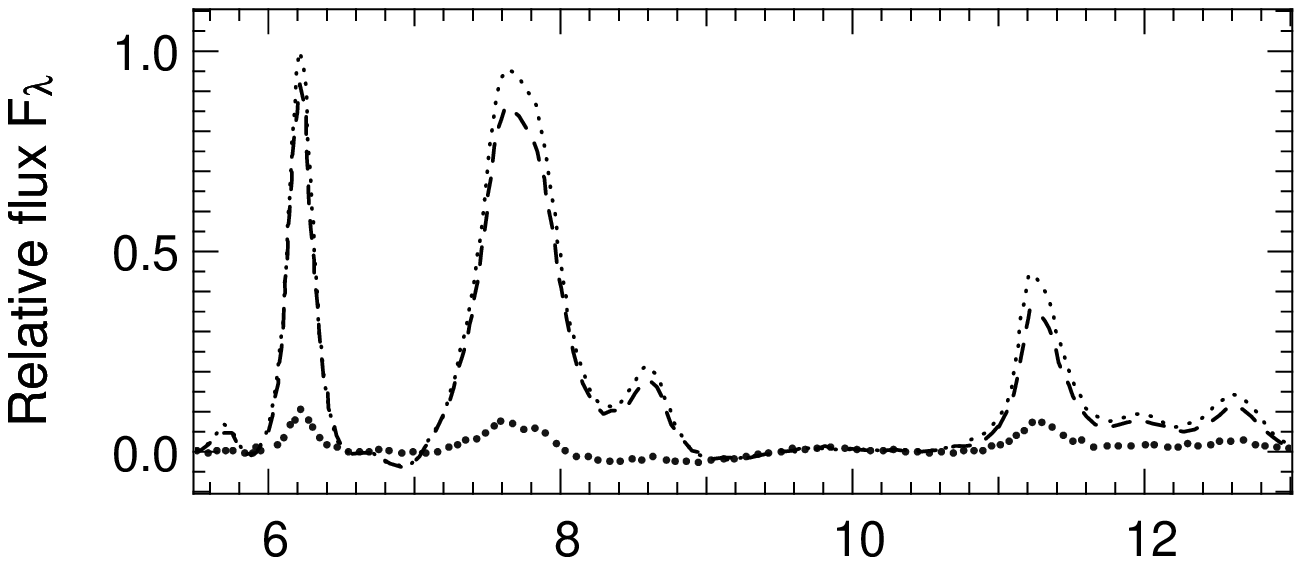}\\
\includegraphics[width=\columnwidth]{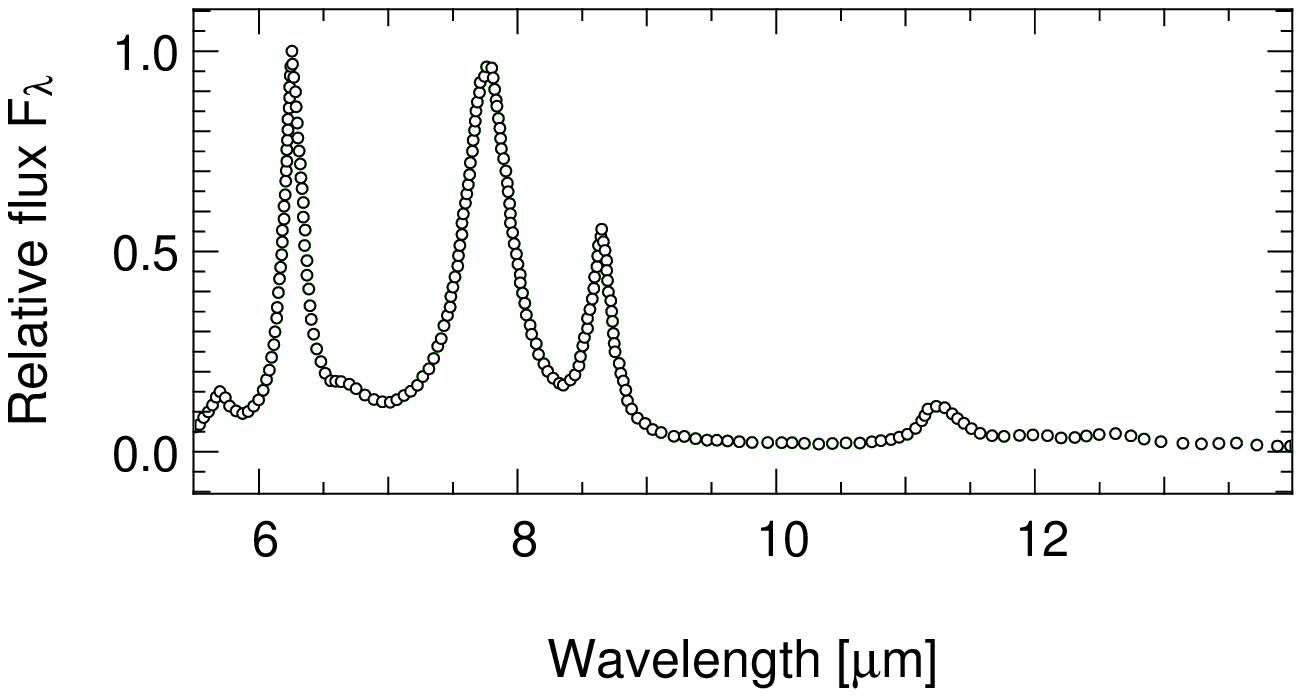}
\caption{PAHs emission (i.e., continuum-subtracted) models for IRS\,48 (top) and HD\,179218 (bottom) adapted from \cite{Maaskant2014} and \cite{Seok2017}, respectively. 
The spectra are normalized to the peak emission at 6.3\,$\mu$m. Top: the small-dotted line corresponds to the total PAH emission, the dashed line to the emission from ionized PAHs in optically thin environments like the gap, and the large-dotted line corresponds to the emission from neutral PAHs in the optically thick disk. Bottom: the model for the total PAH emission in HD\,179218 shows a prominent feature at 8.6\,$\mu$m and a weaker feature at 11.3\,$\mu$m.}\label{pah}
\end{figure}
Our results can be replaced in the larger context of the study of PAH emission in Herbig stars' disks. 
When looking at the question of PAHs spatial extent and charge state (ionized vs. neutral) in more detail, the recent studies by \cite{Maaskant2013,Maaskant2014} highlight more complex scenarios. 
It is found from four typical Herbig objects that the bulk of the PAH emission in (pre)transitional Herbig systems can originate either in the inner (optically thin) or the outer (optically thick) region of the protoplanetary disk. Observationally, this results in a PAH emission component with a smaller, comparable or larger characteristic size than the thermal continuum emission. 
Two representative cases of, respectively, compact and extended emission are seen in IRS\,48 \citep{Geers2007} and HD\,97048 \citep{Lagage2006}. 
Interestingly, \cite{Maaskant2014} suggests a correlation between a) the relative spatial extent of the PAH emission with respect to the thermal continuum and b) the charge state of the PAH molecules as classically traced by diagnostics such as the $I_{6.2}/I_{11.3}$ or $I_{3.3}/I_{7.7}$ feature ratios -- see Fig.~21 in \cite{Peeters2002} --  or the relative strength of the 7.7+8.6\,$\mu$m feature compared to the 11.3\,$\mu$m feature. 
In the archetypical case of HD\,97048, the PAH emission is found to be significantly more extended than the thermal continuum at both 8.6 and 11.3\,$\mu$m \citep{Lagage2006,Doucet2007,Maaskant2013}. At the same time, the object's mid-IR spectroscopy is indicative of an emission caused predominantly by neutral PAH molecules traced by the strong 3.3 and 11.3 features \citep{Seok2017, Maaskant2014}. 
The opposite case is found with IRS\,48 where the ionized state of PAHs, as suggested by a $I_{6.2}/I_{11.3}$ ratio larger than unity, goes together with the 11.3-$\mu$m PAH emission size being more compact than the continuum emission \citep{Maaskant2014}. \\
In HD\,179218, different works based on {\it ISO} and {\it Spitzer} spectroscopy have reported and confirmed the relatively stronger 8.6-$\mu$m PAH feature compared to the 11.3-$\mu$m one \citep{Meeus2001, Acke2010, Juhasz2010, Seok2017}. 
Looking at the continuum-subtracted spectrum of HD\,179218 modeled in \cite{Seok2017} and qualitatively comparing it to the one of IRS\,48 indicates similarities between the two spectra in terms of strength of the 6.2 and 7.7-$\mu$m features compared to the 11.3-$\mu$m feature (see Fig.~\ref{pah}). At first, this could point to a scenario for HD\,179218 similar to IRS\,48 with predominantly ionized PAHs located, for example, in the inner optically thin gap. 
However, our mid-IR imaging results and emission modeling suggest that the PAH contribution is not confined to the inner 10\,au region of HD\,179218, but may extend out to the outer disk regions, where a neutral charge state of the PAHs may be favored.   
A possible explanation to this scenario is that the stellar luminosity of HD\,179218 ($L$=180$L_{\odot}$, \cite{Alecian2013}) being significantly larger than for HD\,97048 ($L$$\sim$40$L_{\odot}$, \cite{Lagage2006,Maaskant2014}), HD\,169142 ($L$$\sim$10$L_{\odot}$, \cite{Alecian2013,Maaskant2014}) or for HD\,135344 ($L$$\sim$10--15$L_{\odot}$, \cite{Alecian2013,Maaskant2014}), the central star consequently produces a stronger UV radiation field capable of ionizing PAH molecules out to larger distances. A plausible hypothesis could also be that PAH ionization out to large distances results from a wide-angle wind impinging on the disk surface. In a different context, this scenario is observed in higher-mass evolved Wolf-Rayet stars \citep{Marchenko2017}. Ideally, the wind scenario could be investigated with IR interferometry by resolving the spatial size of the Br$\gamma$ emission line detected in HD\,179218 \citep{GarciaLopez2006} and comparing it to the size of the nearby continuum, similarly to the case of MWC297, whose Br$\gamma$ emission is driven by a disk-wind mechanism \citep{Malbet2007}. For example, in the case of HD\,97048, recent GRAVITY interferometric observations revealed a Br$\gamma$ emission more compact than the nearby continuum (K. Rousselet-Perraut, private communication) possibly indicative of a magnetospheric accretion process mechanism taking place in the very inner disk region \citep{Kraus2008}. Lacking a clear disk-wind mechanism could hence justify the survival and abundance of neutral PAH in that system. It is particularly interesting to note that, by tracing the ro-vibrational line of molecular H$_{\rm 2}$ at 2.12\,$\mu$m in HD\,97048's disk, \cite{Bary2008} conclude at a quiescent state of H$_{\rm 2}$ that is not shocked nor entrained in a fast-moving wind or outflow associated with this young source. 
 
 \subsection{An alternative disk radiative transfer model}
 
In this work, we also propose an alternative radiative transfer model to \cite{Dominik2003} for the disk of HD\,179218 by fitting simultaneously the SED and our imaging data. Dust thermal emission and isotropic scattering are considered in our radiative transfer model, but PAH emission is not included. Keeping in mind the uncertainty on the object's parallax, our best model implies a larger outer disk than D03 with $R_{\rm o}$=80\,au and, in particular, a negative power-law with $p$=-1.5 for the outer disk surface density. This is slightly lower than the $p$=-1 found for the outer disks of similar group Ia/b objects like HD\,100546 \citep{Tatulli2011} , HD\,139614 \citep{Matter2016} or AB\,Aur \citep{diFolco2009}. This is nonetheless opposite to the positive power-law ($p$=+2) derived by D03, and we find that any positive dust density law in which most of the mass is located in the outer regions of the disk would result in a surface brightness distribution that should be resolved by our observations at 12.5\,$\mu$m.

\section{Conclusions}

We conducted mid-infrared imaging and spectroscopic observations of the Herbig star HD\,179218 using CanariCam on the GTC and obtained the following results.

\begin{itemize}
\item Helped by good weather conditions and by the format of CanariCam images into cubes of short duration savesets, we were able to obtain close to diffraction limited images of HD\,179218, and among the sharpest N-band images obtained from the ground with a FWHMs of $\sim$210\,mas at 8\,$\mu$m. By re-centering and combining a large number of savesets, we reached 3$\sigma$ uncertainties of less than 5\,mas on the FWHM.  With this potential, we resolve for the first time the circumstellar emission around HD\,179218 in the PAH bands at 8.6 and 11.3\,$\mu$m and found characteristic size of $\sim$100\,mas in diameter. \\
\item We performed photometry of the system at 8.6, 11.3 and 12.5\,$\mu$m and found values consistent with published flux densities and without noticeable variability within the measurement errors. The CanariCam low-resolution spectrum matches quite well the shape measured by {\it Spitzer}, 
except in the region of the Earth's ozone band where the spectral calibration is found to be unreliable.\\
\item Importantly, the combination of our imaging data with radiative transfer modeling suggests that the spatially resolved emission at 8.6 and 11.3\,$\mu$m is not of thermal equilibrium nature but may originate from UV-excited PAH molecules located at the surface of the flared disk. 
By taking into account the relative flux ratios between the PAH and thermal component, we find that our observations are best reproduced with a model of PAH "disk" extending out to the physical limits of the dust disk model.\\
\item We discuss the compatibility of such a spatial scenario with the spectroscopic evidence that a predominant fraction of the PAH molecules might be in an ionized charge state. We suggest that a particularly strong UV radiation field from the star or a disk wind may ionize the PAH molecules out to the largest radii. \\
\item In contrast to the disk model already proposed by \cite{Dominik2003}, our alternative radiative transfer model of HD\,179218 coupled to mid-infrared imaging at 12.5\,$\mu$m suggests a surface density with a p=-1.5 negative power-law index, with most of the dust mass located in the first 30\,au of the outer disk.  Assuming that the gas and the PAH molecules are strongly coupled, the detection of an extended PAH emission would point to a flared structure of the disk in HD\,179218 and confirm earlier results. 
\end{itemize}
 
\begin{acknowledgements}
This work is fully supported by the University of Cologne, the Ministry of Higher Education and Scientific Research in Iraq (MoHERS), the Bonn-Cologne Graduate School of Physics and Astronomy (BCGS) and the University of Baghdad. We are thankful to all members of the Gran Telescopio Canarias (GTC) telescope for their support during the observations. We would also like to express our gratitude to the Sky Team of the IAC for providing us with the seeing data. We thank the anonymous referee for assessing the quality of this work.
\end{acknowledgements} 
\bibliographystyle{bibtex/aa}
\bibliography{anas.bib}     

\appendix

\section{Cumulative FWHM}\label{A-appendix}

\begin{figure}
\includegraphics[width=0.9\columnwidth]{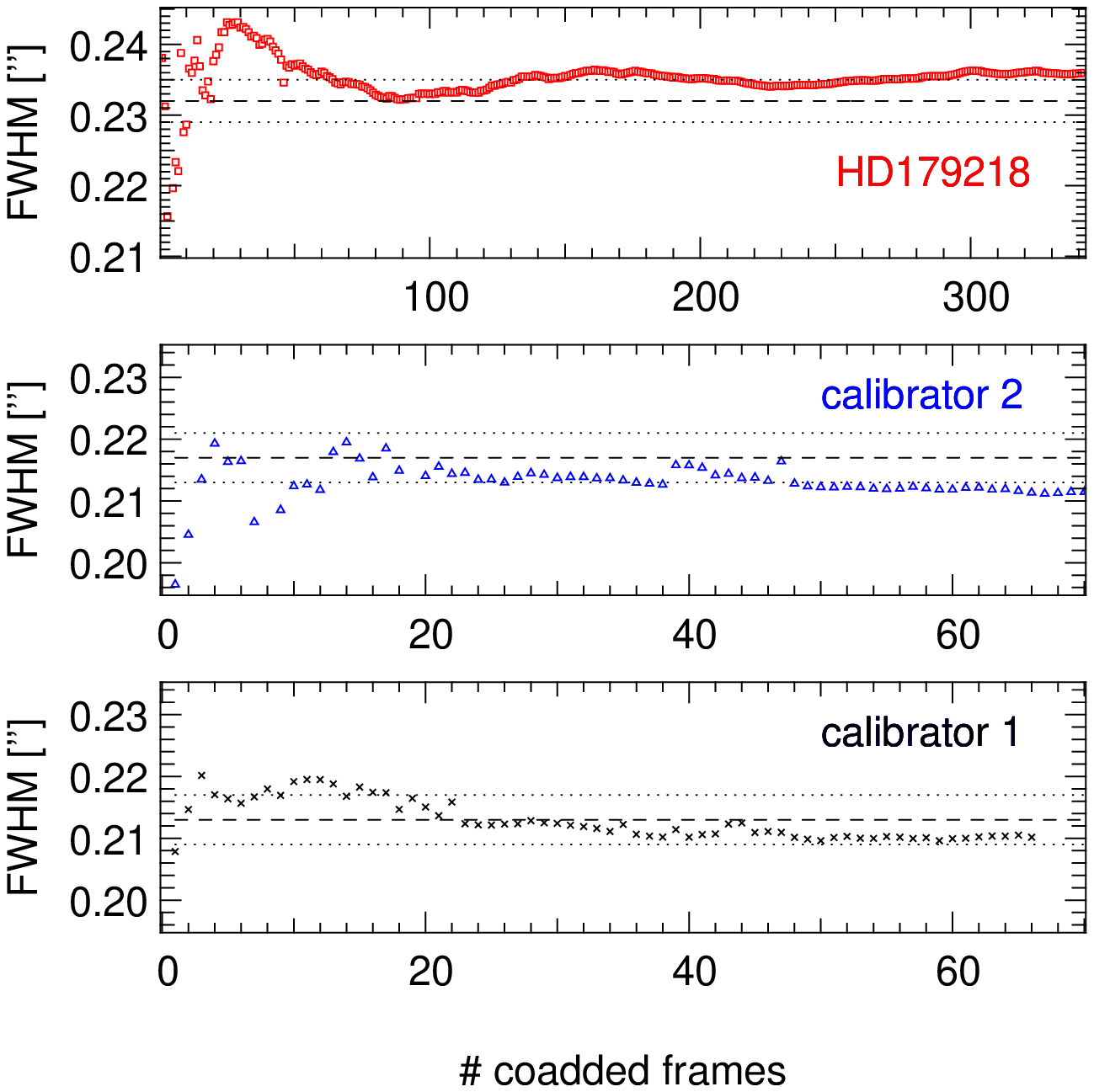}\\
\includegraphics[width=0.9\columnwidth]{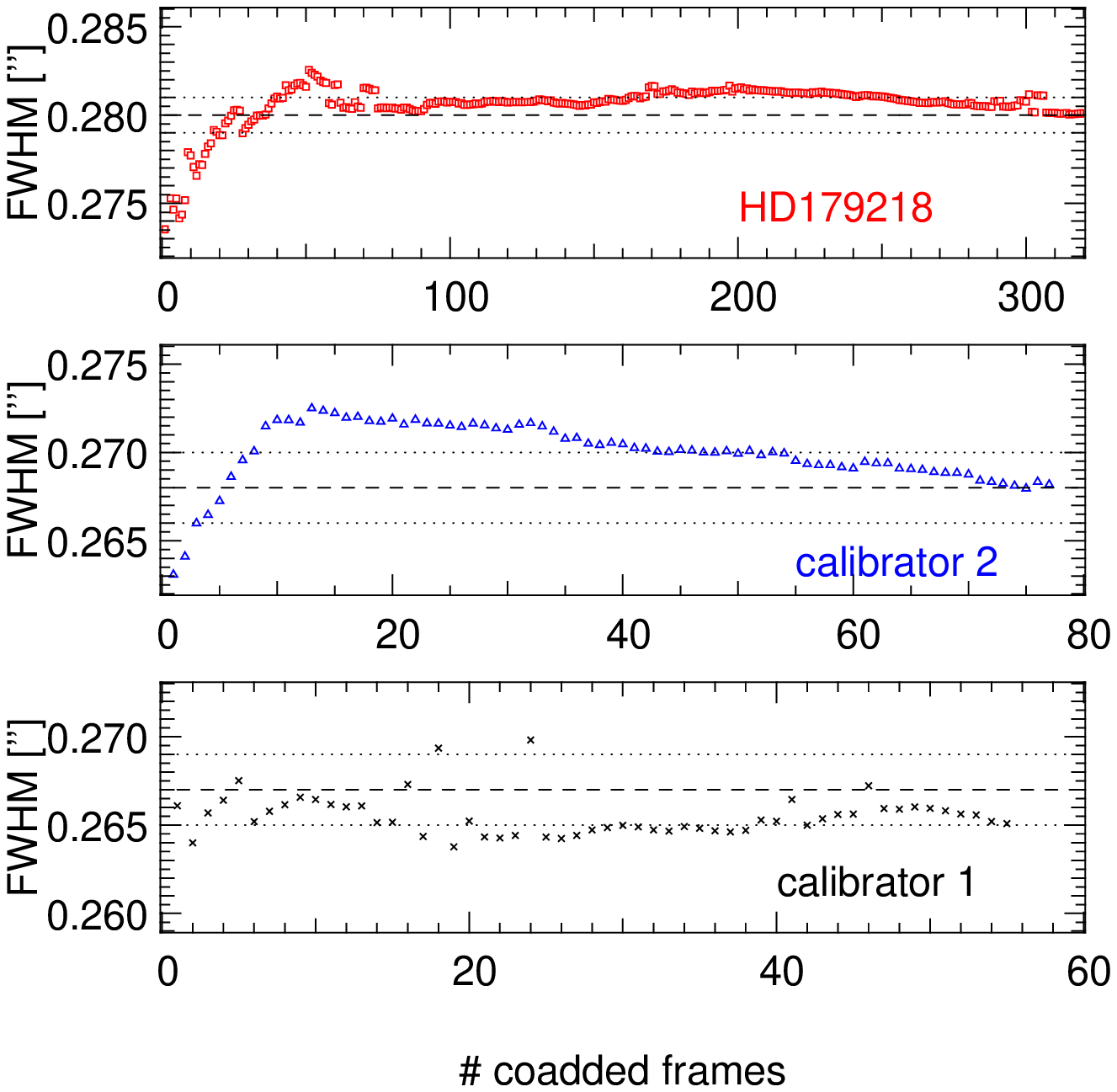}\\
\includegraphics[width=0.9\columnwidth]{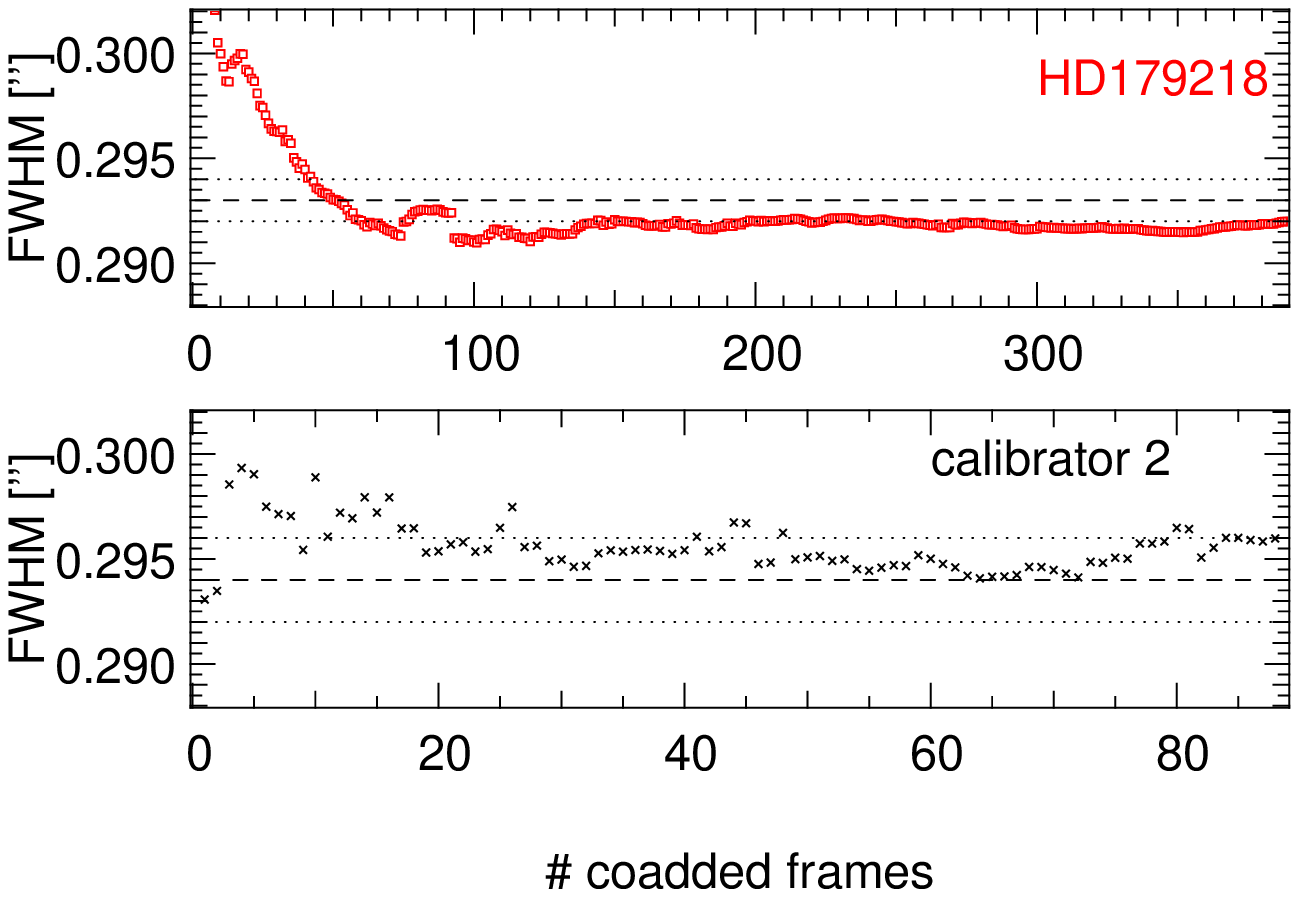}
\caption{Cumulative FWHM for both reference stars (HD\,169414 and HD\,187642) and science star (HD\,179218) in the PAH-1 (top), PAH-2 (center) and Si-6 (bottom) filters as a function of the number of co-added frames following the procedure described in Sect.~\ref{profile}. The dashed line is the average FWHM reported in Table~\ref{tableFWHM} and following a Lorentzian fit, whereas the dotted lines correspond to the 3$\sigma$ boundaries.}\label{Fig:cumulative}
\end{figure}

The following plots in Fig.~\ref{Fig:cumulative} illustrate how the sub-pixel re-centering of the individual savesets with respect to each other allows us to minimize the broadening of the PSF that could result from long integrations or basic stacking. 

\newpage

\section{Results of $\chi^2$ minimization}\label{B-appendix}

\begin{figure}[h]
\includegraphics[trim=4.0cm 11cm 4.0cm 1.5cm, clip, width=\columnwidth, angle=0]{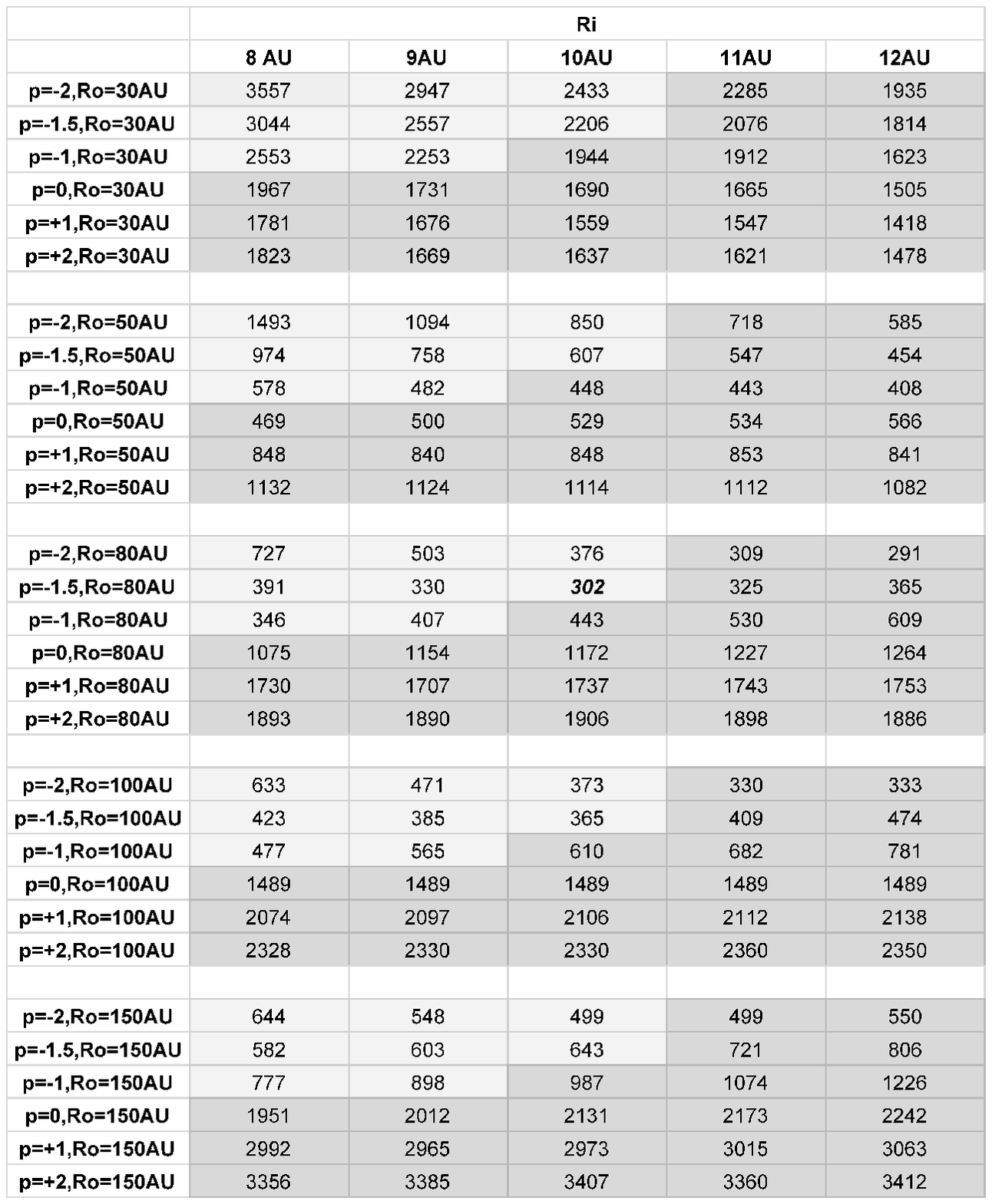}
\caption{Non-reduced $\chi^2$ table for the fit of the SED for different values of $p$ in Eq.~\ref{Eq3}, $R_{\rm i}$ and $R_{\rm o}$, respectively the inner and outer radius of the outer disk in HD\,179218. The light and dark gray boxes correspond to models for which the 12.5\,$\mu$m PSF is spatially unresolved and resolved, respectively. The best model is identified for the value $\chi^2$=302 (bold and italic).}\label{tab:chi2}
\end{figure}

Table~\ref{tab:chi2} reports the value of the non-reduced $\chi^2$ on the SED obtained during the search of our best radiative transfer model of the disk's thermal emission in HD\,179218. The observational SED is plotted in Fig.~\ref{mysed}. We compare a posteriori the synthetic and the observed PSF profiles at 12.5\,$\mu$m.  

\end{document}